\documentclass[reprint,amsmath,amssymb,pra,floatfix,]{revtex4-2}

\usepackage{graphicx}
\usepackage{dcolumn}
\usepackage{bm}
\usepackage{threeparttable}
\usepackage{hyperref}
\usepackage{multirow}
\usepackage{xcolor} 

\usepackage{overpic}

\usepackage{fancyhdr}

\makeatletter

\newcommand{\@aionnum}{xxxxxx}

\fancypagestyle{plain}{%
\fancyhf{} 
\fancyhead[R]{AION-REPORT/\@aionnum}
}

\newcommand{\aionnum}[1]{%
    \def\@aionnum{#1}%
}

\makeatother

\newcommand{\clock}{${}^1S_0 - {}^3P_0$}

\def\RB {Ramsey-Bord\'e}
\def\pp {$\pi$-pulse}
\def\bs {$\pi/2$-pulse}
\def\Max {\mathrm{Max}}
\def\dt {\Delta t_{\mathrm{LMT}}}
\def\dtau {\Delta\tau_{\mathrm{LMT}}}
  
\newcolumntype{e}[1]{D{.}{.}{#1}}

\makeatletter
\AtBeginDocument{
\g@addto@macro{\appendix}{\renewcommand{\p@subsection}{}}
}
\makeatother

\begin{document}

\title{A single-photon large-momentum-transfer atom interferometry scheme for Sr or Yb atoms with application to determining the fine-structure constant}
\aionnum{2024-02}

\author{Jesse S. Schelfhout} \email{jesse.schelfhout@physics.ox.ac.uk}
\affiliation{%
Department of Physics, University of Oxford, Parks Road, Oxford, OX1 3PU, United Kingdom
}%
\author{Thomas M. Hird}
\affiliation{%
Department of Physics, University of Oxford, Parks Road, Oxford, OX1 3PU, United Kingdom
}%
\author{Kenneth M. Hughes}
\affiliation{%
Department of Physics, University of Oxford, Parks Road, Oxford, OX1 3PU, United Kingdom
}%
\author{Christopher J. Foot}
\affiliation{%
Department of Physics, University of Oxford, Parks Road, Oxford, OX1 3PU, United Kingdom
}%

\begin{abstract}
The leading experimental determinations of the fine-structure constant, $\alpha$, currently rely on atomic photon-recoil measurements from \RB\ atom interferometry with large momentum transfer to provide an absolute mass measurement. We propose an experimental scheme for an intermediate-scale differential atom interferometer to measure the photon-recoil of neutral atomic species with a single-photon optical clock transition. We calculate trajectories for our scheme that optimise the recoil phase while nullifying the undesired gravity-gradient phase by considering independently launching two clouds of ultracold atoms with the appropriate initial conditions. For Sr and Yb, we find an atom interferometer of height 3\,m to be sufficient for an absolute mass measurement precision of $\Delta m / m \sim 1\times 10^{-11}$ with current technology. Such a precise measurement (the first of its kind for Sr or Yb) would halve the uncertainty in $\alpha$ --- an uncertainty that would no longer be limited by an absolute mass measurement. The removal of this limitation would allow the uncertainty in $\alpha$ to be reduced by a factor of 10 by corresponding improvements in relative mass measurements, thus paving the way for higher-precision tests of the Standard Model of particle physics.
\end{abstract}

\maketitle
\thispagestyle{plain}

\section{\label{sec:introduction}Introduction}
An atomic wavefunction brought into a superposition of energy eigenstates that freely evolve for some time before being interfered is a precision measurement technique developed by Ramsey \cite{Ramsey1950} that forms the basis of the microwave atomic clocks currently used to realise the SI second \cite{Lombardi2007}. The use of Ramsey pulse sequences for high-precision optical spectroscopy of thermal atoms required the introduction of additional atom-light interactions \cite{Baklanov1976,Bergquist1977,Borde1977,Chebotaev1978}; typically a second set of Ramsey pulses is used that propagates in the opposite direction to the first set \cite{Borde1981,Borde1984}. This technique is a form of atom interferometry, now known as \RB\ interferometry, that results in two closed-loop matter-wave interference paths with a difference in their phase from the atomic recoils along the light propagation axis \cite{Borde1989}. It was found to be important for tests of quantum mechanics, particularly through the fine-structure constant \cite{Weiss1993,Weiss1994,Wicht2002,Gerginov2006,Clade2006,Cadoret2008,Chiow2009,Bouchendira2011,Parker2018,Morel2020}.

The leading experimental values for the fine-structure constant, electron mass, and atomic mass constant \cite{CODATA2022,Tiesinga2021}, hence also many other dependent constants (including $\mu_0,\varepsilon_0,a_0,\mu_B$) and the absolute masses of most particles, rely on atomic photon-recoil measurements from \RB\ interferometry with Rb \cite{Morel2020} and Cs \cite{Parker2018} atoms. The fine-structure constant and electron mass are free parameters in the Standard Model that require empirical determination so theoretical predictions, e.g. the magnetic moment of the electron \cite{Fan2023} or the energy levels of positronium \cite{Adkins2022}, can be evaluated. Accordingly, \RB\ interferometry facilitates the most precise comparison between experiment and theory under the Standard Model \cite{Fan2023}. However, a discrepancy of more than 5$\sigma$ exists between the values from the different atomic species, with the theoretically-derived value lying in between. This discrepancy limits the precision with which the measurements of the anomalous magnetic moment of the electron can test the Standard Model \cite{Fan2023}, motivating additional high-precision determinations of $\alpha$.

The Kasevich-Chu interferometer, based on Raman transitions, demonstrated gravimetry using light-pulse atom interferometry \cite{Kasevich1992}; the simultaneous operation of two such gravimeters in a differential configuration was later used as a gradiometer \cite{Snadden1998}. A gravitational wave detector was proposed using the same principle but with a very long baseline between the interferometers \cite{Dimopoulos2008a,Dimopoulos2009}. An improved scheme using single-photon transitions, in which the laser phase cancels in a differential measurement, was subsequently proposed \cite{Yu2010,Graham2013}. Intermediate-scale ($\sim 10\,\mathrm{m}$) prototype instruments operating on this principle with Sr or Yb atoms are in development \cite{Badurina2020,Abe2021,Hartwig2015}.

In this paper, we investigate the prospect of performing \RB\ interferometry with single-photon transitions in neutral optical clock atoms in intermediate-scale atom interferometers for high-precision determinations of the fine-structure constant. One of the limiting factors in upscaling interferometric measurements of photon recoil is the gravity-gradient phase, which a scheme using offset simultaneous conjugate interferometers has been proposed to address \cite{Zhong2020}. We use this principle to devise an experimental scheme in which two clouds of ultracold atoms are independently launched from their sources in order to mitigate the gravity gradient phase, and large momentum transfer (LMT) is used to enhance the sensitivity to the atomic photon-recoil frequency.

We outline the theoretical framework that we use to derive phase shift expressions in Sec. \ref{sec:theory}, before describing the proposed experimental scheme in Sec. \ref{sec:scheme}. We present the calculation of the differential phase for the proposed scheme in Sec. \ref{sec:differential} and the gravity gradient phase in Sec. \ref{sec:gravity_gradient}. In Sec. \ref{sec:case study}, we determine optimal parameters for the trajectories in two configurations for various instrument sizes. We discuss the determination of the fine-structure constant in Sec. \ref{sec:discussion}, before concluding in Sec. \ref{sec:conclusion}.

\section{\label{sec:theory}Theoretical framework}
The output of an atom interferometer can usually be calculated under the Feynman path integral formalism using the classical action \cite{Storey1994}. The difference in phase between the two arms of an atom interferometer can be decomposed into (i) free propagation (including internal evolution), (ii) separation from closure, and (iii) interaction with the laser \cite{Dimopoulos2008}:
\begin{equation}
    \Delta\phi_{\mathrm{total}} = \Delta\phi_{\mathrm{propagation}} + \Delta\phi_{\mathrm{separation}} + \Delta\phi_{\mathrm{laser}}.\label{eq:d_phi_general}
\end{equation}
For flat spacetime and in the non-relativistic limit, the difference in propagation phase $\int p_{\mu}dx^{\mu} / \hbar$ between the arms is approximated by
\begin{align}\label{eq:d_phi}
    \Delta\phi_{\mathrm{propagation}} &\approx \frac{1}{\hbar} \int \left(\frac{m \Delta (v^2)}{2} - \Delta U - \Delta E \right) dt,
\end{align}
where $\Delta (v^2)$ and $\Delta U$ are the differences in squared velocity and potential energy between the two trajectories, respectively, and $\Delta E$ is the difference in energy eigenvalues between the trajectories. Integration is implicitly assumed to be over the interferometry sequence. Following Overstreet \textit{et al.} \cite{Overstreet2021}, we express the contribution proportional to $\Delta v^2$ in terms of the mean position, $\overline{\mathbf{x}}$, and separation, $\Delta\mathbf{x}$, (and their derivatives) as
\begin{align}
    \frac{m}{\hbar} \int \frac{\Delta(v^2)}{2}dt &= \frac{m}{\hbar} \int \dot{\overline{\mathbf{x}}}\cdot\Delta\dot{\mathbf{x}} dt \nonumber\\
    &= \frac{\overline{\mathbf{p}}(t_f)\cdot\Delta\mathbf{x}(t_f)}{\hbar} - \frac{m}{\hbar} \int \ddot{\overline{\mathbf{x}}}\cdot\Delta\mathbf{x} dt
\end{align}
Since $\Delta\phi_{\mathrm{separation}} = -\overline{\mathbf{p}}(t_f)\cdot\Delta\mathbf{x}(t_f)/\hbar$, we have
\begin{align}
    \Delta\phi_{\mathrm{total}} &=  \Delta\phi_{\mathrm{laser}} - \frac{1}{\hbar} \int \left(m\ddot{\overline{\mathbf{x}}}\cdot\Delta\mathbf{x} + \Delta U + \Delta E \right)dt. \label{eq:dphi_total_intermediate}
\end{align}

The laser phase is defined when the pulse leaves the laser, but in practice we may calculate the laser phase at the points of interaction with an atomic wavepacket relative to some other spacetime coordinate. In absorption of a pulse of laser light, the phase of the laser beam adds to the phase of the atomic wavepacket, whereas the opposite occurs for stimulated emission. This influence of the laser beam phase on that of wavepacket $j$ is taken into account by
\begin{align}
    (\phi_{\mathrm{laser}})_j &= \sum_i \eta_{ji} \Big(\mathbf{k}_{ji}\cdot\mathbf{x}_j(t_{ji}) - \omega_{ji} t_{ji} - \phi_{ji}\Big),
\end{align}
where $\mathbf{k}_{ji}$, $\omega_{ji}$, $\phi_{ji}$, and $t_{ji}$ are respectively the wavevector, (angular) frequency, phase at the reference coordinate, and time of interaction for laser pulse $i$ at wavepacket $j$, and
\begin{align}
    \eta_{ji} &= \begin{cases}
        1, & \mathrm{stimulated\ absorption}\\
        -1, & \mathrm{stimulated\ emission}\\
        0, & \mathrm{no\ interaction}
    \end{cases}
\end{align}
qualifies the interaction of laser pulse $i$ with atomic wavepacket $j$.

The external degrees of freedom of the wavepacket $j$ with trajectory $\mathbf{x}_j(t)$ are governed by $m\ddot{\mathbf{x}}_j(t) + \nabla U(\mathbf{x}_j(t)) = 0$. We denote the free-fall trajectory of the atom in the absence of laser impulses by $\mathbf{x}_0(t)$, and the Green's function for the equation of motion by $\mathbf{G}(t,s)$. Taking the atom-light interaction to provide an instantaneous transfer of momentum $\hbar \eta_{ji} \mathbf{k}_{ji}$ at time $t_{ji}$, the equation of motion is supplemented with a source term $\hbar \eta_{ji} \mathbf{k}_{ji} \delta (t - t_{ji}) / m$ and the particular solution acquires an additional term
\begin{equation}
    \frac{\hbar}{m} \eta_{ji} \mathbf{k}_{ji} \cdot \int_{-\infty}^{\infty} \mathbf{G}(t,s) \delta (s - t_{ji}) ds = \frac{\hbar}{m} \eta_{ji} \mathbf{k}_{ji} \cdot \mathbf{G}(t,t_{ji}).
\end{equation}
We may hence write the classical trajectories of the interferometers as
\begin{align}
    \mathbf{x}_j(t) = \mathbf{x}_0(t) + \frac{\hbar}{m} \sum_i \eta_{ji} \mathbf{k}_{ji} \cdot \mathbf{G}(t,t_{ji}) + \delta \mathbf{x}_j(t), \label{eq:general trajectory}
\end{align}
where $\delta \mathbf{x}_j(t)$ accounts for perturbations to $\mathbf{x}_0(t)$ arising from the spatial variation of the gradient of the potential.

Restricting ourselves to the 1-D case of a vertically-oriented atom interferometer on the surface of the Earth and neglecting gravitational redshift of the laser pulses between the wavepackets, we find $\mathbf{k}_{ji}\equiv\chi_i k_i \hat{\mathbf{z}}$, where
\begin{align}
    \chi_i &= \begin{cases}
        1, & \mathrm{upwards}\\
        -1, & \mathrm{downwards}
    \end{cases}
\end{align}
indicates the direction of laser pulse $i$, $\omega_{ji}\equiv\omega_i$, and $\phi_{ji}\equiv\phi_i$. We note that the product $\eta_{ji} \chi_i$ indicates the direction of momentum transfer (positive indicating upwards) of laser pulse $i$ to atomic wavepacket $j$.

For an interferometer beginning in the ground state comprised of classical trajectories $\mathbf{x}_1(t) \equiv z_1(t) \hat{\mathbf{z}}$ and $\mathbf{x}_2(t) \equiv z_2(t) \hat{\mathbf{z}}$, the terms in the phase shift (\ref{eq:dphi_total_intermediate}) are
\begin{align}
    -\frac{1}{\hbar}\int \Delta E dt &= \omega_0 \sum_i \left[\eta_{1,i} t_{1,i} - \eta_{2,i} t_{2,i}\right]
\end{align}
\begin{align}
    &-\frac{1}{\hbar}\int \left(m\ddot{\overline{\mathbf{x}}}\cdot\Delta\mathbf{x} + \Delta U\right) dt \nonumber\\
    &= - \frac{1}{2}\int \sum_i \chi_i k_i \Big[\eta_{1,i} \delta(t-t_{1,i}) + \eta_{2,i} \delta(t-t_{2,i}) \Big] \Delta z(t) dt \nonumber\\
    &\quad + \Delta\phi_{\mathrm{potential}} \label{eq:propagation_2}
\end{align}
where we follow Overstreet \textit{et al.} \cite{Overstreet2021} in defining 
\begin{align}
    \Delta\phi_{\mathrm{potential}} &= \frac{1}{\hbar}\int \bigg[\left(\frac{\partial U(z_1,t)}{\partial z} + \frac{\partial U(z_2,t)}{\partial z}\right) \frac{\Delta z(t)}{2} \nonumber\\
    &\quad\quad\quad\quad - U(z_1,t) + U(z_2,t)\bigg] dt.
\end{align}
The other term in (\ref{eq:propagation_2}) can be simplified using the integral property of the Dirac delta function that
\begin{equation}
    \int \delta(t - t_{ji}) \Delta z(t) dt = z_1(t_{ji}) - z_2(t_{ji}).
\end{equation}

The difference in laser phase imprinted on the clouds is given by
\begin{align}
    \Delta\phi_{\mathrm{laser}} &= \sum_i \Big[\eta_{1,i}\left(\chi_i k_i z_1(t_{1,i}) - \omega_i t_{1,i} - \phi_i\right) \nonumber\\
    &\quad - \eta_{2,i}\left(\chi_i k_i z_2(t_{2,i}) - \omega_i t_{2,i} - \phi_i\right)\Big].
\end{align}

A number of terms in these expressions are similar and can be grouped together, resulting in a total phase shift of
\begin{align}
    \Delta\phi_{\mathrm{total}} &= \sum_i \chi_i k_i \left[\eta_{1,i} \overline{z}(t_{1,i}) - \eta_{2,i} \overline{z}(t_{2,i}) \right] \nonumber\\
    &\quad + \sum_i (\omega_0 - \omega_i) \left[\eta_{1,i} t_{1,i} - \eta_{2,i} t_{2,i}\right] \nonumber\\
    &\quad + \Delta\phi_{\mathrm{potential}} - \sum_i \phi_i \left[\eta_{1,i} - \eta_{2,i}\right]. \label{eq:general phase shift}
\end{align}
As in Ref. \cite{Overstreet2021}, this expression contains a potential term and the midpoint phase shift. We have included the phase from internal evolution and made explicit that the laser pulses will reach the arms at different times --- to first order, $t_{1,i} - t_{2,i} \approx \chi_i \Delta z(t_{1,i}) / c$ (see Appendix \ref{app:maths.result}).

\section{\label{sec:scheme}Experimental scheme}
\begin{figure}[t]
    \centering
    \includegraphics[width=\columnwidth]{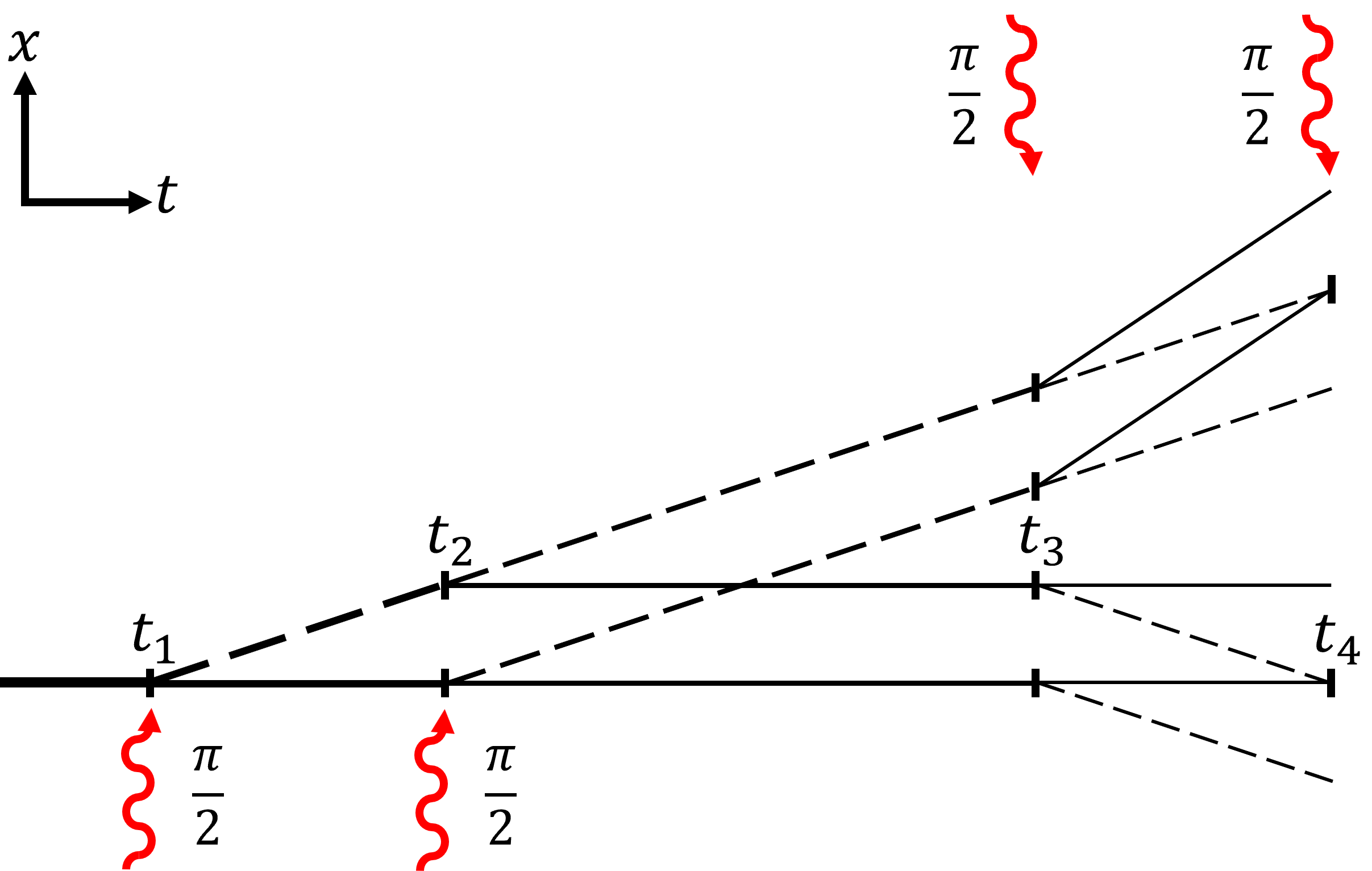}
    \caption{Schematic spacetime diagram of a \RB\ atom interferometer in the absence of external acceleration. The ground (excited) state is indicated by a solid (dashed) line with thickness indicative of the probability density. The timings of the laser pulses are such that $t_4 - t_3 = t_2 - t_1$, thus two pairs of trajectories result in interference.}
    \label{fig:RB_schematic}
\end{figure}

The standard \RB\ atom interferometer consists of four {\bs}s, as shown in Fig.\ \ref{fig:RB_schematic}. The time interval between the first two {\bs}s must be equal to that between the final two for the trajectories to close, and the final two {\bs}s must be in the opposite direction to the first two. This scheme leads to two sets of trajectories that close \footnote{Due to the light propagation time between the trajectories across all of the laser pulses, the interferometers are unlikely to perfectly close} and hence interfere due to the final \bs, as is evident in Fig.\ \ref{fig:RB_schematic}.

The phase difference due to atomic photon-recoil scales as the difference in recoil energy between the arms of each interferometer \cite{Borde1993}. The sensitivity of a \RB\ interferometer can, thus, be enhanced via additional atom-light interactions in two ways. Firstly, using a sequence of $2N$ {\pp}s between the first two and between the final two {\bs}s \cite{Borde1993} to increase the momentum separation from $\hbar k$ to $(N+1)\hbar k$ (and then to decrease it again). This type of scheme has been adopted for large area/large momentum transfer Mach-Zehnder atom interferometry \cite{McGuirk2000,Rudolph2020}; however, for \RB\ interferometry, we wish to transfer momentum to only one arm of each interferometer. Secondly, the two interferometers can be deflected away from each other using a sequence of $M$ \pp s between the second and third \bs s while the arms of each interferometer are in the same atomic state but these states differ between the interferometers \cite{Weiss1993}. Such a scheme is a single-photon transition analogue to the Bragg diffraction and Bloch oscillation scheme used in \cite{Parker2018}.

The differential recoil phase shift between the two interferometers is dependent upon the sequence of atom-light interactions and will persist if the interferometers are compared from two independent cold atom clouds. The ability to control the trajectories of independently-launched interferometers offers the prospect of nullifying the phase arising from first-order spatial variation of the gravitational acceleration, akin to the offset simultaneous conjugate scheme of Zhong \textit{et al.} \cite{Zhong2020}.

Let us consider an idealised 1-D set-up in which all of the laser pulses have perfect fidelity for all atomic wavepackets, have frequency $\omega_i \equiv k_i c = \omega_0 \equiv k c$ that is on resonance with the atoms (neglecting Doppler shifts), and are infinite plane waves so that the momentum transferred to the atoms is equal to the wavevector \cite{Gibble2006}. We will assume the cold atom clouds are sufficiently cold that their thermal expansion is negligible. We will also approximate the gravitational potential near the surface of the Earth to be $U(z) = U_0 + mgz$, where $g>0$, for which $\Delta\phi_{\mathrm{potential}} = 0$. Under these assumptions, (\ref{eq:general phase shift}) becomes
\begin{equation}
    \Delta\phi_{\mathrm{total}} = \sum_i \chi_i k_i \left[\eta_{1,i} \overline{z}(t_{1,i}) - \eta_{2,i} \overline{z}(t_{2,i}) \right] - \phi_i \left[\eta_{1,i} - \eta_{2,i}\right] \label{eq:simple_MP_phase}
\end{equation}
for each interferometer.

As the dynamics are restricted to the vertical and since $U'(z) = mg$ is independent of $z$, thus $\delta \mathbf{x}_j(t) = 0$, the trajectories in (\ref{eq:general trajectory}) can be simplified to 
\begin{equation}
    z_j(t) = z_0(t) + \frac{\hbar k}{m} \sum_i \eta_{ji} \chi_i G(t,t_{ji}), \label{eq:simplified trajectory}
\end{equation}
where $\mathbf{x}_j(t) = z_j(t)\hat{\mathbf{z}}$.

The equation of motion $m \ddot{z}_0(t) + U'(z_0(t)) = 0$ gives $\ddot{z}_0(t) = -g$, for which the Green's function is $G(t,s) = (t-s) \theta(t-s) \equiv \Max\{0,t-s\}$. With initial conditions $z_0(0) = h$ and $\dot{z}_0(0) = u$, the solution is
\begin{align}
    z_0(t) &= h + u t - \frac{g t^2}{2}.
\end{align}
For a different set of initial conditions, $h \rightarrow h + \Delta h$ and $u \rightarrow u + \Delta u$, the solution becomes
\begin{align}
    \widetilde{z_0}(t) &= z_0(t) + \Delta h + \Delta u t.
\end{align}

Neglected higher-order terms in the gravitational potential provide small perturbations to these trajectories; the leading-order contribution comes from the gravity gradient, $G_{zz}$, as $U_{\mathrm{GG}}(z) = m G_{zz} z^2 / 2$. The interferometer phase arising from the gravity gradient shall be calculated perturbatively \cite{Storey1994,Zhong2020} as
\begin{align}
    \Delta\phi_{\mathrm{GG}} &= -\frac{m G_{zz}}{2 \hbar} \int \left(z_1(t)^2 - z_2(t)^2\right) dt \nonumber\\
    &= -\frac{m G_{zz}}{\hbar} \int \overline{z}(t) \Delta z(t) dt. \label{eq:GG}
\end{align}
This phase is calculated in section \ref{sec:gravity_gradient} and conjugate trajectories for which it cancels are presented in section \ref{sec:case study}.

Our proposed scheme involves four classical trajectories, which can be written as
\begin{widetext}
\begin{align}
    z_{\alpha}(t) &= z_0(t) + \zeta(t,t_1) - \zeta(t,t_2) - \zeta(t,t_3) + \sum_{n=1}^N \left[\zeta(t,t_1 + t^{(n)}) - \zeta(t,t_2 - t^{(n)}) - \zeta(t,t_3 + t^{(n)}) + \zeta(t,t_4 - t^{(n)})\right] \nonumber\\
    &\quad - \sum_{m=1}^{M} \zeta(t,t_2 + \tau^{(m)}) \label{eq:z_alpha}\\
    z_{\beta}(t) &= z_0(t) - \sum_{m=1}^{M} \zeta(t,t_2 + \tau^{(m)} + \delta\tau_{\beta}^{(m)}) \label{eq:z_beta}\\
    z_{\gamma}(t) &= \widetilde{z_0}(t) + \zeta(t,t_1 + \Delta t_1) + \sum_{n=1}^N \left[\zeta(t,t_1 + t^{(n)} + \Delta t_1^{(n)}) - \zeta(t,t_2 - t^{(n)} + \Delta t_2^{(n)})\right] + \sum_{m=1}^{M} \zeta(t,t_2 + \tau^{(m)} + \delta\tau_{\gamma}^{(m)}) \label{eq:z_gamma}\\
    z_{\delta}(t) &= \widetilde{z_0}(t) + \zeta(t,t_2 + \Delta t_2) + \zeta(t,t_3 + \Delta t_3) + \sum_{n=1}^N \left[\zeta(t,t_3 + t^{(n)} + \Delta t_3^{(n)}) - \zeta(t,t_4 - t^{(n)} + \Delta t_4^{(n)})\right] \nonumber\\
    &\quad + \sum_{m=1}^{M} \zeta(t,t_2 + \tau^{(m)} + \delta\tau_{\delta}^{(m)}), \label{eq:z_delta}
\end{align}
\end{widetext}
where the impulse response function $\zeta(t,t_i)$ encodes the times at which momentum kicks occur, and is defined by
\begin{align}
    \zeta(t,t_i) &= \frac{\hbar k}{m} G(t,t_i) \equiv v_r \Max\{0,t - t_i\}, \label{eq:zeta}
\end{align}
where $v_r \equiv \hbar k / m$ is the atomic recoil velocity due to an impulse at time $t_i$. These expressions for the trajectories are applicable from launch until the final \bs .

The timing of the laser pulses is defined by their intersection with the trajectory $z_{\alpha}$, giving rise to quantities $\Delta t_i$ ($i\in\{1,2,3,4\}$), $\Delta t_i^{(n)}$ ($i\in\{1,2,3,4\},\ n\in\{1,...,N\}$), and $\delta\tau_{j}^{(m)}$  ($j\in\{\beta,\gamma,\delta\},\ m\in\{1,...,M\}$) that account for light propagation between trajectories.

\begin{figure}[t]
    \centering
    \def\svgwidth{\columnwidth}
    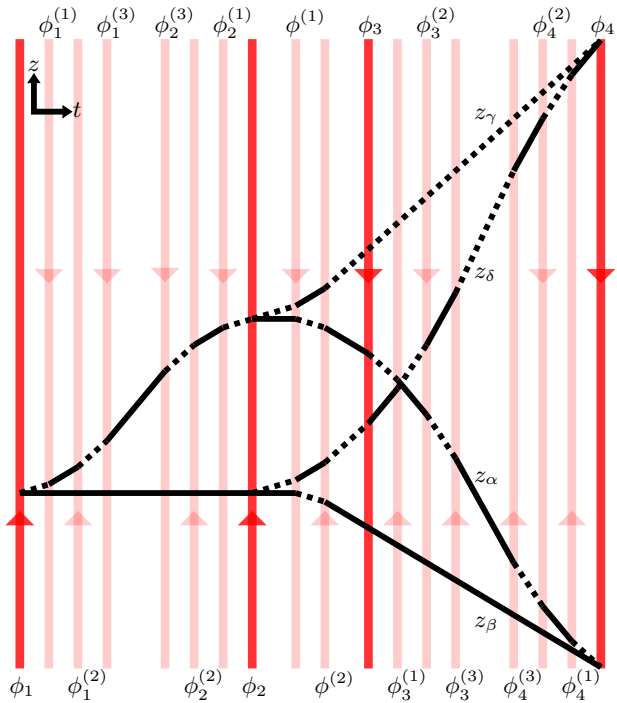
    \caption{Schematic spacetime diagram of an enhanced \RB\ atom interferometer with $N=3$ additional recoil-energy-difference pulses and $M=2$ deflecting pulses. Only the paths leading to closed interferometers are shown, indicated by solid (dotted) lines for the ground (excited) state. The \bs s (\pp s) are represented by strong (faint) red vertical lines with k-vectors in the directions indicated by the arrowheads. Each laser pulse is ascribed a phase symbol $\phi_i$ that is used to identify its interaction parameters with each trajectory according to Table \ref{tab:all_pulses}. The path labels $z_j$ correspond to trajectories in the free-fall frame with $\Delta h = 0 = \Delta u$.}
    \label{fig:Enhanced_RB}
\end{figure}

A spacetime diagram of the trajectories in the free-fall frame for the same initial conditions (i.e. $\Delta h = 0 = \Delta u$) is presented in Fig. \ref{fig:Enhanced_RB}. We partition the laser pulses into three groups: (1) \RB\ pulses, (2) additional recoil-energy-difference (ARED) pulses, and (3) deflecting pulses.

The \RB\ pulses are the \bs s labelled by $\phi_i$ in Fig. \ref{fig:Enhanced_RB} that are also seen in Fig. \ref{fig:RB_schematic}. They have directions $\chi_{\{1,2\}} = 1$ (both are upwards) and $\chi_{\{3,4\}} = (-1)^{M+1}$ (e.g. both downwards for $M=2$).

The ARED pulses are pairs of velocity-selective \pp s that increase and later decrease the recoil-energy difference between the arms of each interferometer. They are accounted for by the sums over $n$ in (\ref{eq:z_alpha}) -- (\ref{eq:z_delta}) and are labelled by $\phi_i^{(n)}$ in Fig. \ref{fig:Enhanced_RB}. One pulse reaches $z_{\alpha}$ at time $t_1 + t^{(n)}$ or $t_3 + t^{(n)}$ and its partner at time $T - 2 t^{(n)}$ later. Both pulses in each pair have the same direction and adjacent pairs are directed oppositely. The proposed scheme has these pairs oriented such that $\chi^{(n)} = (-1)^n$ in the interval $(t_1,t_2)$, and $\chi^{(n)} = (-1)^{M+n+1}$ in the interval $(t_3,t_4)$. One pulse in the pair stimulates absorption, the other emission, so that the net momentum transfer vanishes after both pulses, but the additional momentum between the pulses results in displacement.

The deflecting pulses are \pp s that interact with all trajectories, alternating between stimulating absorption in one direction and emission in the other. They are accounted for by the sums over $m$ in (\ref{eq:z_alpha}) -- (\ref{eq:z_delta}) and are labelled by $\phi^{(m)}$ in Fig. \ref{fig:Enhanced_RB}. Our scheme has these pulses oriented such that $\chi^{(m)} = (-1)^m$ in the interval $(t_2,t_3)$.

Our trajectories form two interferometers, one with $z_{\alpha}$ and $z_{\beta}$ and one with $z_{\gamma}$ and $z_{\delta}$. The deflecting pulses kick the trajectories $z_{\alpha}$ and $z_{\beta}$ downwards and so we shall call their interferometer the `diminished interferometer'. Likewise, $z_{\gamma}$ and $z_{\delta}$ are kicked upwards so their interferometer shall be called the `augmented interferometer'.

\subsection{\label{subsec:diminished}Diminished interferometer}
The mean position, $\overline{z}_d$, and separation, $\Delta z_d = z_{\alpha} - z_{\beta}$, between the arms of the diminished interferometer are given by
\begin{widetext}
\begin{align}
    \overline{z}_d(t) &= z_0(t) + \frac{1}{2} \left(\zeta(t,t_1) - \zeta(t,t_2) - \zeta(t,t_3) + \sum_{n=1}^N \left[\zeta(t,t_1 + t^{(n)}) - \zeta(t,t_2 - t^{(n)}) - \zeta(t,t_3 + t^{(n)}) + \zeta(t,t_4 - t^{(n)})\right] \right.\nonumber\\
    &\quad\left. - \sum_{m=1}^{M} \left[\zeta(t,t_2 + \tau^{(m)}) + \zeta(t,t_2 + \tau^{(m)} + \delta\tau_{\beta}^{(m)})\right]\right) \label{eq:z_diminished}\\
    \Delta z_d(t) &= \zeta(t,t_1) - \zeta(t,t_2) - \zeta(t,t_3) + \sum_{n=1}^N \left[\zeta(t,t_1 + t^{(n)}) - \zeta(t,t_2 - t^{(n)}) - \zeta(t,t_3 + t^{(n)}) + \zeta(t,t_4 - t^{(n)})\right] \nonumber\\
    &\quad - \sum_{m=1}^{M} \left[\zeta(t,t_2 + \tau^{(m)}) - \zeta(t,t_2 + \tau^{(m)} + \delta\tau_{\beta}^{(m)})\right]. \label{eq:dz_diminished}
\end{align}
\end{widetext}
All of the laser pulses interact with $z_{\alpha}$, but only the deflecting pulses interact with $z_{\beta}$ (except for the final \bs\ for one of the output ports --- we will use the other output port in our calculation). For the ARED pulses, we have $\eta^{(n)} = (-1)^n$ for the first pulse in the pair and $\eta^{(n)} = (-1)^{n+1}$ for the second. For the deflecting pulses, $\eta^{(m)} = (-1)^{m+1}$ and so $\chi^{(m)} \eta^{(m)} = -1$ in the interval $(t_2,t_3)$, i.e. the arms are deflected downwards.

\subsection{\label{subsec:augmented}Augmented interferometer}
The mean position, $\overline{z}_a$, and separation, $\Delta z_a = z_{\gamma} - z_{\delta}$, between the arms of the augmented interferometer are given by
\begin{widetext}
\begin{align}
    \overline{z}_a(t) &= \widetilde{z_0}(t) + \frac{1}{2} \Bigg(\zeta(t,t_1 + \Delta t_1) + \zeta(t,t_2 + \Delta t_2) + \zeta(t,t_3 + \Delta t_3) \nonumber\\
    &\quad\left. + \sum_{n=1}^N \left[\zeta(t,t_1 + t^{(n)} + \Delta t_1^{(n)}) - \zeta(t,t_2 - t^{(n)} + \Delta t_2^{(n)}) + \zeta(t,t_3 + t^{(n)} + \Delta t_3^{(n)}) - \zeta(t,t_4 - t^{(n)} + \Delta t^{(n)})\right] \right.\nonumber\\
    &\quad\left. + \sum_{m=1}^{M} \left[\zeta(t,t_2 + \tau^{(m)} + \delta\tau_{\gamma}^{(m)}) + \zeta(t,t_2 + \tau^{(m)} + \delta\tau_{\delta}^{(m)})\right]\right) \label{eq:z_augmented}\\
    \Delta z_a(t) &= \zeta(t,t_1 + \Delta t_1) - \zeta(t,t_2 + \Delta t_2) - \zeta(t,t_3 + \Delta t_3) + \sum_{m=1}^{M} \left[\zeta(t,t_2 + \tau^{(m)} + \delta\tau_{\gamma}^{(m)}) - \zeta(t,t_2 + \tau^{(m)} + \delta\tau_{\delta}^{(m)})\right] \nonumber\\
    &\quad + \sum_{n=1}^N \left[\zeta(t,t_1 + t^{(n)} + \Delta t_1^{(n)}) - \zeta(t,t_2 - t^{(n)} + \Delta t_2^{(n)}) - \zeta(t,t_3 + t^{(n)} + \Delta t_3^{(n)}) + \zeta(t,t_4 - t^{(n)} + \Delta t_4^{(n)})\right] . \label{eq:dz_augmented}
\end{align}
\end{widetext}
All of the laser pulses before $t_2$ and the deflecting pulses interact with $z_{\gamma}$, while all of the pulses from $t_2$ onwards interact with $z_{\delta}$. Like the diminished interferometer, the final \bs\ interacts with a different trajectory for each output port to produce the same phase difference --- we will use the other output port that interacts with $z_{\delta}$ in our calculation. For the ARED pulses, in the interval $(t_1,t_2)$ we have $\eta^{(n)} = (-1)^n$ for the first pulse in the pair and $\eta^{(n)} = (-1)^{n+1}$ for the second; the reverse is true for the interval $(t_3,t_4)$. For the deflecting pulses, $\eta^{(m)} = (-1)^m$ and so the product $\chi^{(m)} \eta^{(m)} = 1$ in the interval $(t_2,t_3)$, i.e. the interferometer is deflected upwards.

\section{\label{sec:differential}Differential phase}
The differential phase can be found by modifying (\ref{eq:simple_MP_phase}) to include both the augmented and diminished interferometers. We find
\begin{align}
    \Delta\Phi &= \Delta\phi_a - \Delta\phi_d \nonumber\\
    &= \sum_i \bigg[\chi_i k \Big(\eta_{\gamma i} \overline{z}_a(t_{\gamma i}) - \eta_{\delta i} \overline{z}_a(t_{\delta i}) - \eta_{\alpha i} \overline{z}_d(t_{\alpha i}) \nonumber\\
    &\quad + \eta_{\beta i} \overline{z}_d(t_{\beta i})\Big) + \phi_i\left(\eta_{\gamma i} - \eta_{\delta i} - \eta_{\alpha i} + \eta_{\beta i}\right)\bigg]. \label{eq:DPhi}
\end{align}
The laser pulse parameters needed to evaluate (\ref{eq:DPhi}) are displayed in Table \ref{tab:all_pulses}. For all pulses, the sum $\eta_{\gamma i} - \eta_{\delta i} - \eta_{\alpha i} + \eta_{\beta i} = 0$ and so the terms with $\phi_i$ do not contribute to the differential phase.

\begin{table*}[t]
    \centering
    \caption{Configuration of atom-light interactions for differential atom interferometry scheme. Laser pulses may be distinguished by the symbol for their phase given in the first column.}
    \label{tab:all_pulses}
    \begin{ruledtabular}
    \begin{tabular}{lD{1}{1}{2.7}D{1}{1}{1.4}lD{1}{1}{2.7}lD{1}{1}{2.7}lD{1}{1}{2.5}l}
        $\phi_i$ & \multicolumn{1}{D{.}{.}{2.4}}{\chi_i} & \multicolumn{1}{D{.}{.}{1.1}}{\eta_{\gamma i}} & $t_{\gamma i}$ & \multicolumn{1}{D{.}{.}{2.4}}{\eta_{\delta i}} & $t_{\delta i}$ & \multicolumn{1}{D{.}{.}{2.4}}{\eta_{\alpha i}} & $t_{\alpha i}$ & \multicolumn{1}{D{.}{.}{2.1}}{\eta_{\beta i}} & $t_{\beta i}$ \\\hline
        $\phi_1$ & 1 & 1 & $t_1 + \Delta t_1$ & \multicolumn{1}{D{0}{0}{7}}{0} & --- & 1 & $t_1$ & \multicolumn{1}{D{0}{0}{5}}{0} & ---\\
        $\phi_2$ & 1 & \multicolumn{1}{D{0}{0}{4}}{0} & --- & 1 & $t_2 + \Delta t_2$ & -1 & $t_2$ & \multicolumn{1}{D{0}{0}{5}}{0} & ---\\
        $\phi_3$ & (-1)^{M+1} & \multicolumn{1}{D{0}{0}{4}}{0} & --- & (-1)^{M+1} & $t_3 + \Delta t_3$ & (-1)^M & $t_3$ & \multicolumn{1}{D{0}{0}{5}}{0} & ---\\
        $\phi_4$ & (-1)^{M+1} & \multicolumn{1}{D{0}{0}{4}}{0} & --- & (-1)^M & $t_4 + \Delta t_4$ & (-1)^{M+1} & $t_4$ & \multicolumn{1}{D{0}{0}{5}}{0} & ---\\
        $\phi_1^{(n)}$ & (-1)^n & (-1)^n & $t_1 + t^{(n)} + \Delta t_1^{(n)}$ & \multicolumn{1}{D{0}{0}{7}}{0} & --- & (-1)^n & $t_1 + t^{(n)}$ & \multicolumn{1}{D{0}{0}{5}}{0} & ---\\
        $\phi_2^{(n)}$ & (-1)^n & (-1)^{n+1} & $t_2 - t^{(n)} + \Delta t_2^{(n)}$ & \multicolumn{1}{D{0}{0}{7}}{0} & --- & (-1)^{n+1} & $t_2 - t^{(n)}$ & \multicolumn{1}{D{0}{0}{5}}{0} & ---\\
        $\phi_3^{(n)}$ & (-1)^{M+n+1} & \multicolumn{1}{D{0}{0}{4}}{0} & --- & (-1)^{M+n+1} & $t_3 + t^{(n)} + \Delta t_3^{(n)}$ & (-1)^{M+n} & $t_3 + t^{(n)}$ & \multicolumn{1}{D{0}{0}{5}}{0} & ---\\
        $\phi_4^{(n)}$ & (-1)^{M+n+1} & \multicolumn{1}{D{0}{0}{4}}{0} & --- & (-1)^{M+n} & $t_4 - t^{(n)} + \Delta t_4^{(n)}$ & (-1)^{M+n+1} & $t_4 - t^{(n)}$ & \multicolumn{1}{D{0}{0}{5}}{0} & ---\\
        $\phi^{(m)}$ & (-1)^m & (-1)^m & $t_2 + \tau^{(m)} + \delta\tau_{\gamma}^{(m)}$ & (-1)^m & $t_2 + \tau^{(m)} + \delta\tau_{\delta}^{(m)}$ & (-1)^{m+1} & $t_2 + \tau^{(m)}$ & (-1)^{m+1} & $t_2 + \tau^{(m)} + \delta\tau_{\beta}^{(m)}$
    \end{tabular}
    \end{ruledtabular}
\end{table*}

Using the fact that 
\begin{equation}
    \overline{z}_a(t) = z_{\gamma}(t) - \Delta z_a(t) / 2 = z_{\delta}(t) + \Delta z_a(t) / 2 \label{eq:midpoint_za}
\end{equation}
and 
\begin{equation}
    \overline{z}_d(t) = z_{\alpha}(t) - \Delta z_d(t) / 2, \label{eq:midpoint_zd}
\end{equation}
along with the expression (\ref{eq:dt_definition}) for $\Delta t_i$, $\Delta t_i^{(n)}$, and $\delta\tau_j^{(m)}$, we find
\begin{widetext}
\begin{align}
    \Delta\Phi &= k c \left(\Delta t_1 - \Delta t_2 + (-1)^M \left(\Delta t_3 - \Delta t_4\right) + \sum_{n=1}^N (-1)^n \left(\Delta t_1^{(n)} - \Delta t_2^{(n)} + (-1)^M \left(\Delta t_3^{(n)} - \Delta t_4^{(n)}\right) \right) \right.\nonumber\\
    &\quad\left. + \sum_{m=1}^M (-1)^m\left(\delta\tau_{\gamma}^{(m)} - \delta\tau_{\delta}^{(m)} - \delta\tau_{\beta}^{(m)}\right)\right) + \omega_r \Bigg(- 2 (N+1) (N+2M+2) T + (2M+1) \left(\Delta t_1 - \Delta t_2\right) \nonumber\\
    &\quad + \Delta t_3 - \Delta t_4 + \sum_{n=1}^N \Bigg[4 (2M+n+2) t^{(n)} + (2M-n+1) \left(\Delta t_1^{(n)} - \Delta t_2^{(n)}\right) + (n+1) \left(\Delta t_3^{(n)} - \Delta t_4^{(n)}\right) \nonumber\\
    &\quad + \sum_{p=n}^N \left[4 t^{(p)} + \Delta t_2^{(p)} - \Delta t_1^{(p)} + \Delta t_3^{(p)} - \Delta t_4^{(p)}\right] \Bigg] \Bigg), \label{eq:derivation_result_main_text}
\end{align}
\end{widetext}
where $\omega_r = \hbar k^2/(2m) \equiv k v_r/2$ is the atomic photon-recoil frequency. A derivation of this result is presented in Appendix \ref{app:b}. Evaluation of this phase difference requires knowledge of the propagation time delays $\Delta t_i$, $\Delta t_i^{(n)}$, and $\delta\tau_j^{(m)}$, and the LMT pulse sequence timings $t^{(n)}$ and $\tau^{(m)}$. We use a pulse sequence that is simple to describe mathematically for the purpose of this calculation:
\begin{align}
    t^{(n)} &= n \dt \label{eq:t^n}\\
    \tau^{(m)} &= t_d - t_2 + (m-1) \dtau, \label{eq:tau^m}
\end{align}
where $t_d$ is the time of the first deflecting pulse.

We arrive at
\begin{align}
    \Delta\Phi &= \frac{(N+1) (N + 2 M + 2) \hbar k^2 (T - N \dt)}{m} \nonumber\\
    &\quad - \frac{N (N+1) (N + 2) \hbar k^2 \dt}{3 m} + O\left(\frac{1}{c}\right) \label{eq:dPhi}
\end{align}

The full expression for (\ref{eq:dPhi}) to order $1/c$ is given by (\ref{eq:dPhi_full}), and its terms are indexed and deconstructed in Table \ref{tab:breakdown}.

The configuration of LMT pulses constrains either the pulse timings or the LMT orders according to
\begin{align}
    (2 N + 1) \dt &\leq T \label{eq:dt_inequality}\\
    M \dtau &\leq t_3 - t_d \label{eq:dtau_inequality},
\end{align}
leading to a lower bound of
\begin{align}
    \Delta\Phi &\gtrsim Q \frac{\hbar k^2 \dt}{m},
\end{align}
where $Q = 2M(N+1)^2 + (N+1)(N+2)(2N+3)/3$ is an (even) integer.
Accordingly, we find the scaling of the recoil frequency times $T$ remains quadratic in $N$ and $M$ (when considering that $(N+1)\dt \sim T/2$). Quadratic scaling of the \RB\ phase with LMT order was proposed by Bord\'e \cite{Borde1993} and has been demonstrated experimentally \cite{PlotkinSwing2018}, and offers an advantage compared to the linear scaling with Bloch oscillation order (with small magnification for Bragg beamsplitters) in the Rb and Cs experiments \cite{Morel2020,Parker2018}.

\section{\label{sec:gravity_gradient}Gravity gradient mitigation}
The phase shift from the gravity gradient can be found by treating the gravity gradient as a perturbation and integrating its effect over the unperturbed trajectories \cite{Storey1994} to arrive at (\ref{eq:GG}). The differential phase between our two interferometers is, hence,
\begin{equation}
    \Delta\Phi_{\mathrm{GG}} = - \frac{m G_{zz}}{\hbar} \int_{t_0}^{t_f} \left(\overline{z}_a(t) \Delta z_a(t)  - \overline{z}_d(t) \Delta z_d(t) \right) dt. \label{eq:differential_gravity_gradient_phase}
\end{equation}

To leading order, the effects of the finite speed of light can be neglected and so $\Delta z_a(t) \approx \Delta z_d(t) \approx \Delta z(t)$, hence,
\begin{align}
    \Delta\Phi_{\mathrm{GG}} &\approx - \frac{m G_{zz}}{\hbar} \int_{t_0}^{t_f} \left(\overline{z}_a(t) - \overline{z}_d(t)\right) \Delta z(t) dt, \label{eq:differential_gravity_gradient_phase_approx_1}
\end{align}
where
\begin{widetext}
\begin{align}
    \Delta z(t) &= \zeta(t,t_1) - \zeta(t,t_2) - \zeta(t,t_3) + \sum_{i=1}^N \left[\zeta(t,t_1 + t^{(i)}) - \zeta(t,t_2 - t^{(i)}) - \zeta(t,t_3 + t^{(i)}) + \zeta(t,t_4 - t^{(i)})\right] \label{eq:dz}\\
    \overline{z}_a(t) - \overline{z}_d(t) &\approx \Delta h + \Delta u t + \zeta(t,t_2) + \zeta(t,t_3) + \sum_{i=1}^N \left[\zeta(t,t_3 + t^{(i)}) - \zeta(t,t_4 - t^{(i)})\right] + 2 \sum_{i=1}^M \zeta(t,t_2 + \tau^{(i)}). \label{eq:zp-zd}
\end{align}

Using (\ref{eq:t^n}) and (\ref{eq:tau^m}) and the linearity of integrals, we arrive at
\begin{align}
    \Delta\Phi_{\mathrm{GG}} &\approx -\frac{(N+1) k G_{zz}}{180} \bigg(180 (T - N \dt) (t_3 - t_1) \Delta h + 45 (T - N \dt) (t_3 - t_1) \Delta u (t_1 + t_2 + t_3 + t_4) \nonumber\\
    &\quad\quad + \frac{15 \hbar k (T - N \dt)}{m} \Big[2 (N + 2 M + 2) T^2 - T (4 N + 2 M + 5) N \dt + 6 (t_3 - t_2) (t_3 - t_1) \nonumber\\
    &\quad\quad\quad + 6 M (t_3 - t_d) (t_4 - (t_d + (M-1) \dtau)) + 6 M (t_4 - t_d) (t_3 - (t_d + (M-1) \dtau)) \nonumber\\
    &\quad\quad\quad + (N+1) (3 N + 2 M + 3) N \dt^2 + 2 M (M-1) (2 M - 1) \dtau^2 \Big]\nonumber\\
    &\quad\quad - \frac{N (N+2) (3 N^2 + 6 N + 1) \hbar k \dt^3}{m} \bigg). \label{eq:dPhi_GG}
\end{align}
\end{widetext}
A derivation of this result is presented in Appendix \ref{app:c}. The terms in (\ref{eq:dPhi_GG}) are indexed and broken down in Table \ref{tab:breakdown_GG}. For a given set of laser pulses, this differential phase can be mitigated by judicious choice of $\Delta h$ and $\Delta u$ by means of asynchronous and/or asymmetric launches.

\begin{table*}[t]
    \centering
    \caption{Breakdown of (\ref{eq:dPhi_GG}) into its component terms, indexed by their position in the expression. The dimensionless products are all either unity or a product of $G_{zz}$, which has dimensions of $\mathrm{T}^{-2}$, and two quantities with dimensions of time.}
    \label{tab:breakdown_GG}
    \begin{ruledtabular}
    \begin{tabular}{llllD{/}{/}{29.3}}
        Index & Angular frequency & Time & Dimensionless product & \multicolumn{1}{D{.}{.}{30.0}}{\text{Numerical coefficient}} \\ \hline
        1 & $k G_{zz} (t_3 - t_1) \Delta h $ & $(T - N \Delta t_{\mathrm{LMT}} )$ & 1 & -(N+1)\\
        2 & $k \Delta u$ & $(T - N \Delta t_{\mathrm{LMT}} )$ & $G_{zz} (t_3 - t_1) (t_1 + t_2 + t_3 + t_4) $ & -(N+1)/4 \\ 
        3 & $k v_r$ & $(T - N \Delta t_{\mathrm{LMT}})$ & $G_{zz} T^2$ & -(N+1)(N+2M+2)/6\\
        4 & $k v_r$ & $(T - N \Delta t_{\mathrm{LMT}})$ & $G_{zz} T \dt$ & N (N+1) (4N+2M+5)/12\\
        5 & $k v_r$ & $(T - N \Delta t_{\mathrm{LMT}})$ & $G_{zz} (t_3 - t_2)(t_3 - t_1)$ & -(N+1)/2\\
        6 & $k v_r$ & $(T - N \Delta t_{\mathrm{LMT}})$ & $G_{zz} (t_3 - t_d) (t_4 - (t_d + (M-1) \dtau))$ & -(N+1)M/2\\
        7 & $k v_r$ & $(T - N \Delta t_{\mathrm{LMT}})$ & $G_{zz} (t_4 - t_d) (t_3 - (t_d + (M-1) \dtau))$ & -(N+1)M/2\\
        8 & $k v_r$ & $(T - N \Delta t_{\mathrm{LMT}})$ & $G_{zz} \dt^2$ & -N (N+1)^2 (3N + 2M + 3)/12\\
        9 & $k v_r$ & $(T - N \Delta t_{\mathrm{LMT}})$ & $G_{zz} \dtau^2$ & -(N+1) M (M-1) (2M-1)/6\\ 
        10 & $k v_r$ & $\dt$ & $G_{zz} \dt^2$ & N (N+1) (N + 2) (3N^2 + 6N + 1)/180
    \end{tabular}
    \end{ruledtabular}
\end{table*}

\section{\label{sec:case study}Case study: Strontium and Ytterbium}
In this case study, the interferometers are assumed to use neutral optical clock atoms and operate on the \clock\ clock transition. The excited state lifetimes, recoil velocities, and recoil frequencies for Sr and Yb are presented in Table \ref{tab:clock_values}. Although there are two fermionic isotopes of Yb, we focus on ${}^{171}\mathrm{Yb}$ and assume the analysis applies equally to ${}^{173}\mathrm{Yb}$ because the differences in the recoil parameters are only a few percent (mostly the difference in atomic masses).

\begin{table}[t]
    \centering
    \caption{Natural lifetime, $\tau$; wavelength, $\lambda$; recoil velocity, $v_r = \hbar k / m$; and recoil frequency, $\omega_r = \hbar k^2 / 2m$, of the \clock\ clock transition in Sr and Yb. The values for $\lambda$ have been truncated at seven significant figures, while $v_r$ and $\omega_r$ have been rounded to five significant figures --- uncertainties are significantly below the presented precision.}
    \label{tab:clock_values}
    \begin{ruledtabular}
    \begin{tabular}{le{4.0}e{3.6}e{1.4}e{5.0}}
        Species & \multicolumn{1}{c}{$\tau\ (\mathrm{s})$} & \multicolumn{1}{c}{$\lambda\ (\mathrm{nm})$} & \multicolumn{1}{c}{$v_r\ (\mathrm{mm}\,\mathrm{s}^{-1})$} & \multicolumn{1}{c}{$\omega_r\ (\mathrm{rad}\,\mathrm{s}^{-1})$} \\\hline
        ${}^{87}\mathrm{Sr}$ & 151\footnotemark[1] & 698.4457... & 6.5737 & 29568\\
        ${}^{171}\mathrm{Yb}$ & 23\footnotemark[2] & 578.4195... & 4.0358 & 21920\\
        ${}^{173}\mathrm{Yb}$ & 26\footnotemark[2] & 578.4209... & 3.9891 & 21666
    \end{tabular}
    \end{ruledtabular}
    \footnotetext[1]{Ref. \cite{Lu2024}}
    \footnotetext[2]{Ref. \cite{Porsev2004}}
\end{table}

The broader $5s^2\,{}^1S_0 - 5s5p\,{}^3P_1$ intercombination line of Sr has been used to demonstrate a record momentum separation of $601\hbar k$ \cite{Wilkason2022thesis} using LMT with Floquet shaped pulses, which bodes well for the $1000\hbar k$ design goal of intermediate-scale vertical atom interferometers \cite{Badurina2020,Abe2021}. With an anticipated Rabi frequency of a few kHz for Sr \cite{Abe2021}, we construct trajectories using pulse separation times of $\dt = \dtau = 1\,\mathrm{ms}$. We note that the larger transition matrix element of the Yb clock transition that results in its shorter natural lifetime would result in a higher Rabi frequency as compared to Sr for the same laser power and beam size.

The requirement to image the interferometers simultaneously leads to the condition that $\{z_{\beta}(t_f), z_{\gamma}(t_f)\} = \{0,L\}$, with whichever trajectory is launched(/dropped) from the cold atom source at $z=L$ ending up at $z=0$ and vice versa. The trajectories are also constrained by the interferometry sequence occurring within a suitable region. The expression in (\ref{eq:dPhi_GG}) is used to enforce $\Delta\Phi_{\mathrm{GG}} = 0$. There is a trade-off between the total number of pulses, $4(N+1)+M$, that can be fitted into a given measurement time (limited by the time of flight in the apparatus) and the phase sensitivity, which is quadratic in $N$ and $M$ to leading order, thus establishing a constrained optimisation problem. The trajectories that optimise the sensitivity to photon-recoil frequency from (\ref{eq:dPhi}) (ignoring the terms of order $1/c$) were found using a numerical search (of type branch and bound) with computations performed in \textsc{mathematica} and \textsc{python}.

In Sec. \ref{subsec:X}, we analyse a configuration in which the top trajectory is dropped from stationary, leading to trajectories that look like an `X'. We also analyse a configuration in which the top trajectory is launched vertically, as in an atomic fountain, in Sec. \ref{subsec:Fountain}.

\subsection{\label{subsec:X}`X' configuration}
We restrict interferometry to occurring in $[0.2\,\mathrm{m},L-0.2\,\mathrm{m}]$ to avoid potential disturbances that may occur close to the atom sources (where the cold atoms are prepared). In order to maximise the total time available for the interferometry sequence, the cloud originating from $z=L$ should be dropped (i.e. released from stationary) and should be the one to receive additional upwards momentum kicks (i.e. $z_{\gamma,\delta}$). Allowing this cloud to drop to a height $z_0\in(0.2\,\mathrm{m},L-0.2\,\mathrm{m}]$, we have
\begin{align}
    h + \Delta h &= z_0 \label{eq:X_1}\\
    u + \Delta u &= -\sqrt{2 g (L - z_0)}. \label{eq:X_2}
\end{align}
Assuming that we are limited by time constraints, we take the pulse separations to be uniformly $\dt=1\,\mathrm{ms}$, leading to pulse timings (taking $t_1=0$)
\begin{align}
    t_2 &= (2N+1)\dt = T\\
    t_d &= 2(N+1)\dt\\
    t_3 &= (2(N+1)+M)\dt\\
    t_4 &= (4N+M+3)\dt.
\end{align}
The final trajectories (after all LMT pulses) are given by
\begin{widetext}
\begin{align}
    z_{\beta}(t) &= h + (4N+M+3) v_r M \dt / 2 + (u - M v_r) t - g t^2 / 2 \label{eq:z_beta_after_LMT}\\
    z_{\gamma}(t) &= h + \Delta h + N^2 v_r \dt - (4N+M+3) v_r M \dt / 2 + (u + \Delta u + (M+1) v_r) t - g t^2 / 2. \label{eq:z_gamma_after_LMT}
\end{align}
The difference between these is
\begin{align}
     \Delta s(t) \equiv z_{\gamma}(t) - z_{\beta}(t) &= \Delta h + N^2 v_r \dt - (4N+M+3) v_r M \dt + (\Delta u + (2M+1) v_r) t \label{eq:ds_X}.
\end{align}
The condition $z_{\gamma}(t_f) = 0$ gives
\begin{equation}
    t_f = \frac{u + \Delta u + (M+1) v_r + \sqrt{(u + \Delta u + (M+1) v_r)^2 + 2 g (h + \Delta h) + (2N^2 - M (4N+M+3)) g v_r \dt }}{g} \label{eq:tf_X},
\end{equation}
\end{widetext}
and the condition $z_{\beta}(t_f) = L$ gives
\begin{equation}
    \Delta s(t_f) = -L \label{eq:X_3}
\end{equation}
for $\Delta s(t)$ defined by (\ref{eq:ds_X}) and $t_f$ by (\ref{eq:tf_X}).

The constraint $\Delta\Phi_{\mathrm{GG}} = 0$ can be solved simultaneously with (\ref{eq:X_1}), (\ref{eq:X_2}) \& (\ref{eq:X_3}) to yield $\{h,\Delta h,u,\Delta u\}$. A numerical search for the maximum value of (\ref{eq:dPhi}) subject to $h,z_{\beta}(t_4) \in [0.2\,\mathrm{m},L-0.2\,\mathrm{m}]$ was conducted for $N,M \in \mathbb{N}_0$ and $z_0 \in (0.2\,\mathrm{m},L-0.2\,\mathrm{m}]$ with a resolution of $1\,\mathrm{mm}$ in $z_0$. The optimal parameters for Sr are presented in Table \ref{tab:Sr_X} and for Yb in Table \ref{tab:Yb_X}. For the cases in which multiple values of $z_0$ gave identical values for $N,M,\Delta\Phi$, we present the median value of $z_0$.

\begin{table*}[ht]
    \centering
    \caption{Optimal parameters for the X configuration with Sr at different values of $L$}
    \label{tab:Sr_X}
    \begin{ruledtabular}
    \begin{tabular}{e{3.1}e{3.0}e{3.0}e{2.3}e{2.5}e{2.5}e{1.7}}
        \multicolumn{1}{c}{$L\ (\mathrm{m})$} & \multicolumn{1}{c}{$N$} & \multicolumn{1}{c}{$M$} & \multicolumn{1}{c}{$h-0.2\,\mathrm{m}\ (\mathrm{mm})$} & \multicolumn{1}{c}{$u\ (\mathrm{m}\,\mathrm{s}^{-1})$} & \multicolumn{1}{c}{$z_0\ (\mathrm{m})$} & \multicolumn{1}{c}{$\Delta\Phi\ (\mathrm{rad})$} \\\hline
        1 & 25 & 50 & 1.260 & 4.80847 & 0.734 & 4.7\times 10^6\\
        1.5 & 43 & 74 & 24.607 & 5.77004 & 1.2145\footnotemark[1] & 2.0\times 10^7\\
        2 & 58 & 99 & 24.350 & 6.67269 & 1.7015\footnotemark[1] & 4.9\times 10^7\\
        2.5 & 71 & 123 & 27.625 & 7.46262 & 2.1955\footnotemark[1] & 9.0\times 10^7\\
        3 & 85 & 136 & 15.210 & 8.18718 & 2.688\footnotemark[1] & 1.4\times 10^8\\
        3.5 & 94 & 165 & 4.041 & 8.89710 & 3.1865\footnotemark[1] & 2.1\times 10^8\\
        4 & 107 & 172 & 3.230 & 9.48367 & 3.6845\footnotemark[1] & 2.9\times 10^8\\
        4.5 & 116 & 192 & 1.656 & 10.0789 & 4.185 & 3.7\times 10^8\\
        5 & 129 & 192 & 6.601 & 10.5753 & 4.684\footnotemark[1] & 4.7\times 10^8\\
        6 & 146 & 221 & 7.483 & 11.6079 & 5.684\footnotemark[1] & 6.9\times 10^8\\
        7 & 168 & 223 & 1.065 & 12.4837 & 6.683 & 9.5\times 10^8\\
        8 & 184 & 241 & 3.726 & 13.3481 & 7.682 & 1.2\times 10^9\\
        9 & 200 & 254 & 0.536 & 14.1531 & 8.680 & 1.5\times 10^9\\
        10 & 215 & 266 & 3.596 & 14.9098 & 9.6775\footnotemark[1] & 1.9\times 10^9\\
        100 & 817 & 748 & 6.890 & 46.9516 & 99.150 & 8.1\times 10^{10}
    \end{tabular}
    \end{ruledtabular}
    \footnotetext[1]{Median value}
\end{table*}

\begin{table*}[ht]
    \centering
    \caption{Optimal parameters for the X configuration with Yb at different values of $L$}
    \label{tab:Yb_X}
    \begin{ruledtabular}
    \begin{tabular}{e{3.1}e{3.0}e{3.0}e{3.3}e{2.5}e{2.5}e{1.7}}
        \multicolumn{1}{c}{$L\ (\mathrm{m})$} & \multicolumn{1}{c}{$N$} & \multicolumn{1}{c}{$M$} & \multicolumn{1}{c}{$h-0.2\,\mathrm{m}\ (\mathrm{mm})$} & \multicolumn{1}{c}{$u\ (\mathrm{m}\,\mathrm{s}^{-1})$} & \multicolumn{1}{c}{$z_0\ (\mathrm{m})$} & \multicolumn{1}{c}{$\Delta\Phi\ (\mathrm{rad})$} \\\hline
        1 & 26 & 43 & 3.320 & 4.77864 & 0.7295\footnotemark[1] & 3.4\times 10^6\\
        1.5 & 42 & 73 & 51.245 & 5.62916 & 1.2135\footnotemark[1] & 1.4\times 10^7\\
        2 & 59 & 87 & 109.608 & 6.33768 & 1.710\footnotemark[1] & 3.4\times 10^7\\
        2.5 & 69 & 122 & 130.432 & 7.11780 & 2.2035\footnotemark[1] & 6.3\times 10^7\\
        3 & 84 & 129 & 166.491 & 7.74969 & 2.701\footnotemark[1] & 1.0\times 10^8\\
        3.5 & 95 & 148 & 179.277 & 8.39875 & 3.197 & 1.5\times 10^8\\
        4 & 104 & 171 & 174.444 & 9.04051 & 3.691\footnotemark[1] & 2.0\times 10^8\\
        4.5 & 115 & 182 & 180.980 & 9.60267 & 4.1875\footnotemark[1] & 2.6\times 10^8\\
        5 & 123 & 203 & 162.849 & 10.1867 & 4.682\footnotemark[1] & 3.3\times 10^8\\
        6 & 140 & 233 & 134.815 & 11.2372 & 5.6735\footnotemark[1] & 4.9\times 10^8\\
        7 & 157 & 255 & 102.822 & 12.1906 & 6.666\footnotemark[1] & 6.7\times 10^8\\
        8 & 173 & 275 & 64.493 & 13.0786 & 7.6595\footnotemark[1] & 8.9\times 10^8\\
        9 & 189 & 290 & 27.233 & 13.9005 & 8.6545\footnotemark[1] & 1.1\times 10^9\\
        10 & 201 & 317 & 7.257 & 14.6852 & 9.6545\footnotemark[1] & 1.4\times 10^9\\
        100 & 803 & 903 & 2.722 & 46.2060 & 99.438 & 6.6\times 10^{10}
    \end{tabular}
    \end{ruledtabular}
    \footnotetext[1]{Median value}
\end{table*}

\begin{figure}[t]
    \centering
    \includegraphics[width=\columnwidth]{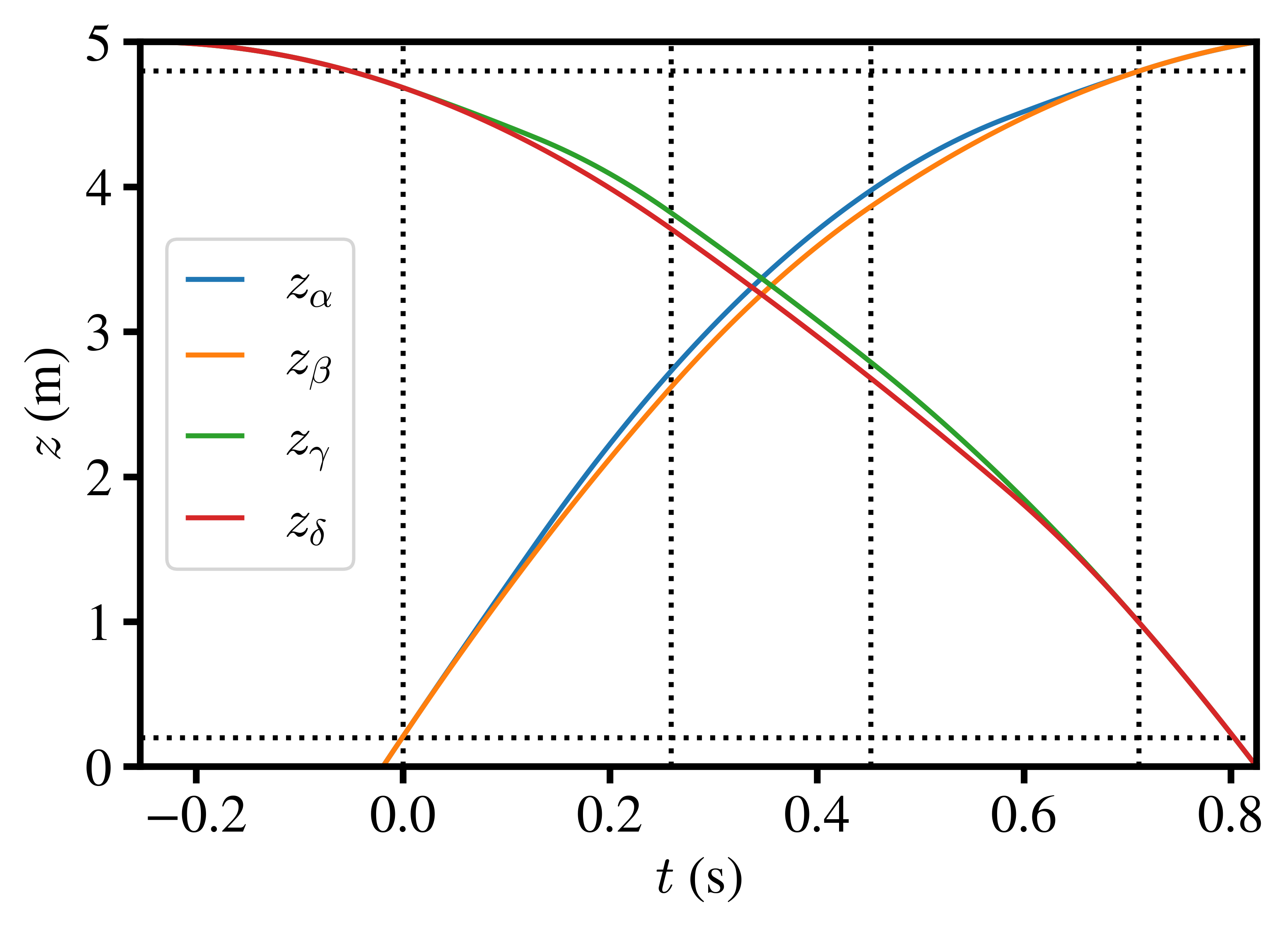}
    \caption{Optimal trajectories from inception for the `X' configuration with $L=5\,\mathrm{m}$ for Sr. The horizontal dotted lines bound the interferometry region, [0.2\,m, 4.8\,m], in space; the vertical dotted lines represent the \bs s and, thus, the outer lines bound the interferometry sequence in time.}
    \label{fig:Sr_X_5m_from_launch}
\end{figure}

The trajectories can be traced back to their times of launch/release from their origins via $t_{\mathrm{top}} = (u + \Delta u)/g \equiv -\sqrt{2 g (L-z_0)}/g$ and $t_{\mathrm{bottom}} = (u - \sqrt{2 g h + u^2})/g$, and the launch velocity of the bottom cloud is given by $u_{\mathrm{launch}} = u - g t_{\mathrm{bottom}}$. Example trajectories traced back to their origin are presented in Fig. \ref{fig:Sr_X_5m_from_launch} for the case $L=5\,\mathrm{m}$ for Sr.

\subsection{\label{subsec:Fountain}Fountain configuration}
We analyse a situation in which the interferometry region is an entire tower of height $L$ with cold atom sources located at $z=0$ and $z=L/2$. The cloud launched from $z=0$ is helped to mitigate the gravity gradient phase by receiving additional upwards momentum kicks (i.e. $z_{\gamma,\delta}$) while the cloud launched from $z=L/2$ receives downwards momentum kicks (i.e. $z_{\alpha,\beta}$). Interferometry time is maximised if the sequence begins when launching the cloud from $z=L/2$, i.e. $h=L/2$.

We consider the general situation in which $T \geq (2N+1) \dt$, but, since $\Delta\Phi$ does not depend on $t_3 - t_2$, we maximise $T$ as a fraction of total time by setting $t_3 - t_2 = (M+1) \dt$. The pulse timings are thus (taking $t_1 = 0$)
\begin{align}
    t_2 &= T\\
    t_d &= T + \dt\\
    t_3 &= T + (M+1)\dt\\
    t_4 &= 2 T + (M+1)\dt.
\end{align}
The final trajectories (after all LMT pulses) are given by
\begin{widetext}
\begin{align}
    z_{\beta}(t) &= h + M v_r (T + (M+1) \dt / 2) + (u - M v_r) t - g t^2 / 2\\
    z_{\gamma}(t) &= h + \Delta h + N v_r (T - (N+1) \dt) - M v_r (T + (M+1) \dt / 2) + (u + \Delta u + (M+1) v_r) t - g t^2 / 2.
\end{align}
The difference between these is
\begin{align}
     \Delta s(t) \equiv z_{\gamma}(t) - z_{\beta}(t) &= \Delta h + N v_r (T - (N+1) \dt) - M v_r (2 T + (M+1) \dt) + (\Delta u + (2M+1) v_r) t \label{eq:ds_Fountain}.
\end{align}
For the time-limited case $T\rightarrow(2N+1)\dt$, these equations reproduce (\ref{eq:z_beta_after_LMT}), (\ref{eq:z_gamma_after_LMT}), and (\ref{eq:ds_X}).

The condition $z_{\beta}(t_f) = 0$ gives
\begin{equation}
    t_f = \frac{u - M v_r + \sqrt{(u - M v_r)^2 + g (2 h + M v_r (2 T + (M+1) \dt))}}{g} \label{eq:tf_Fountain}.
\end{equation}
We may allow the maximum time for interferometry by setting $t_4 = t_f$, which results in
\begin{equation}
    u = \frac{1}{2} \left(M v_r - \frac{2 h}{2 T + (M+1) \dt} + g (2 T + (M+1) \dt)\right).
\end{equation}
\end{widetext}
The condition $z_{\gamma}(t_f) = L/2$ gives
\begin{equation}
    \Delta s(t_f) = L/2 \label{eq:full_1}
\end{equation}
for $\Delta s$ defined by (\ref{eq:ds_Fountain}) and $t_f$ by (\ref{eq:tf_Fountain}).

With $h$ and $u$ determined, the constraint $\Delta\Phi_{\mathrm{GG}} = 0$ can be solved simultaneously with (\ref{eq:full_1}) to yield $\{\Delta h,\Delta u\}$. A numerical search for the maximum value of (\ref{eq:dPhi}) subject to $h+\Delta h,\mathrm{Max}\{z_{\alpha}(t),z_{\gamma}(t)\} \in [0,L]$ and $T\geq (2N+1)\dt$ was conducted for $N,M \in \mathbb{N}_0$.

\begin{figure}[t]
    \centering
    \includegraphics[width=\columnwidth]{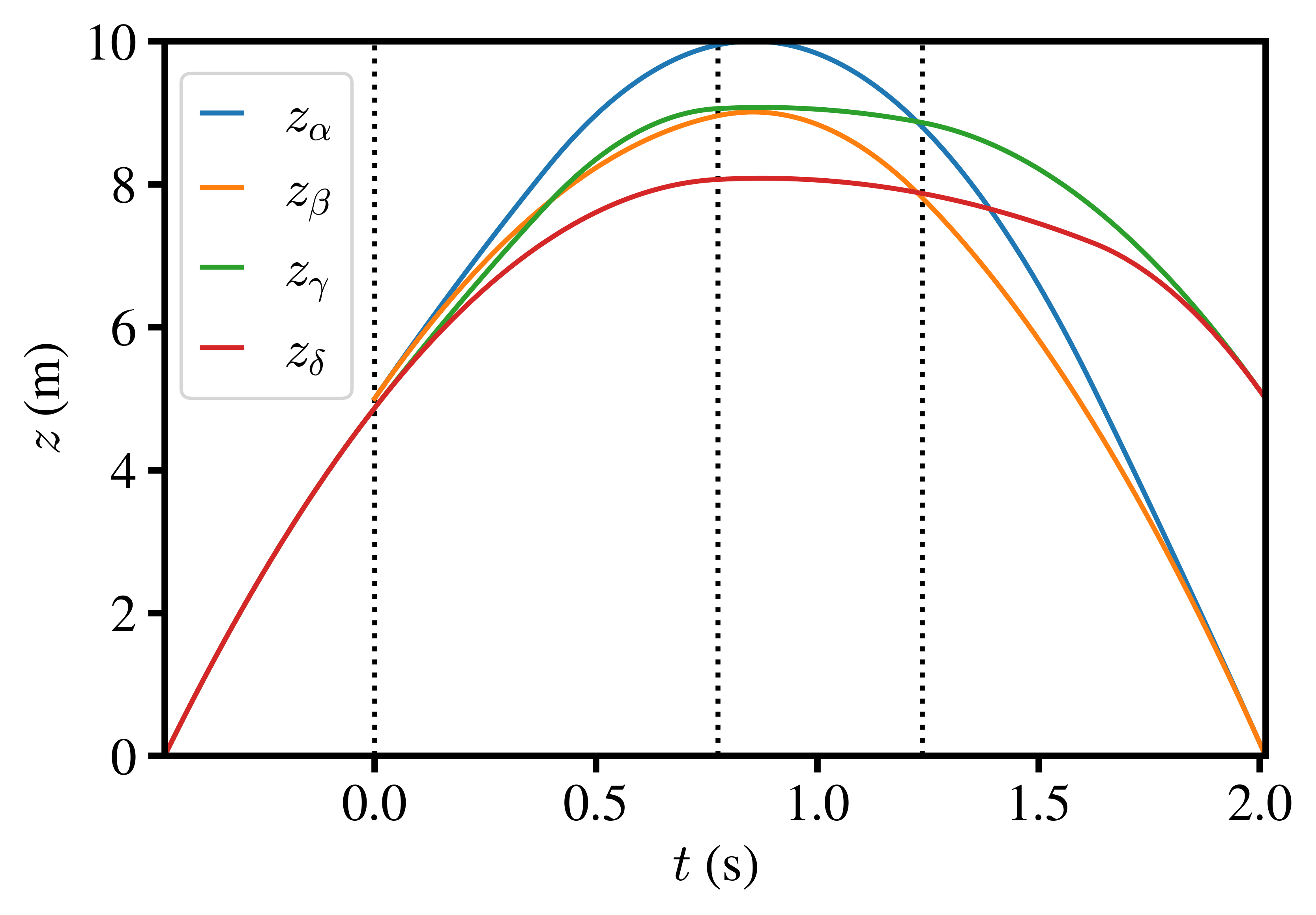}
    \caption{Optimal trajectories from inception for the fountain configuration with $L=10\,\mathrm{m}$ for Sr. The dotted vertical lines indicate the timings of the four \bs s (the final one being coincident with the final time).}
    \label{fig:fountain_from_launch}
\end{figure}

Making $T$ a free parameter necessitates finding the maxima of the trajectories and this increases the computational time for the calculations. Therefore we restricted our search to $L=10\,\mathrm{m}$ for Sr. The optimal parameters were found to be $N=387, M=461, T=0.7756\,\mathrm{s}$, leading to a photon-recoil phase of $\Delta\Phi\approx 1.1\times 10^{10}\,\mathrm{rad}$ --- a factor of 5 larger than the `X' configuration for the same $L$. The corresponding trajectories, traced back to their inception, are presented in Fig. \ref{fig:fountain_from_launch}. The initial conditions of the interferometers for these trajectories are $h = 5\,\mathrm{m}$, $u = 8.906\,\mathrm{m}\,\mathrm{s}^{-1}$, $h + \Delta h = 4.8661\,\mathrm{m}$ and $u + \Delta u = 7.930\,\mathrm{m}\,\mathrm{s}^{-1}$, corresponding to launching the cloud at $z=0\,\mathrm{m}$ with a velocity of $u_{\mathrm{launch}} = 12.584\,\mathrm{m}\,\mathrm{s}^{-1}$ at $t_{\mathrm{launch}} = -474\,\mathrm{ms}$.

\section{\label{sec:discussion}Determination of the fine-structure constant}
\begin{figure}[t]
    \centering
    \includegraphics[width=\columnwidth]{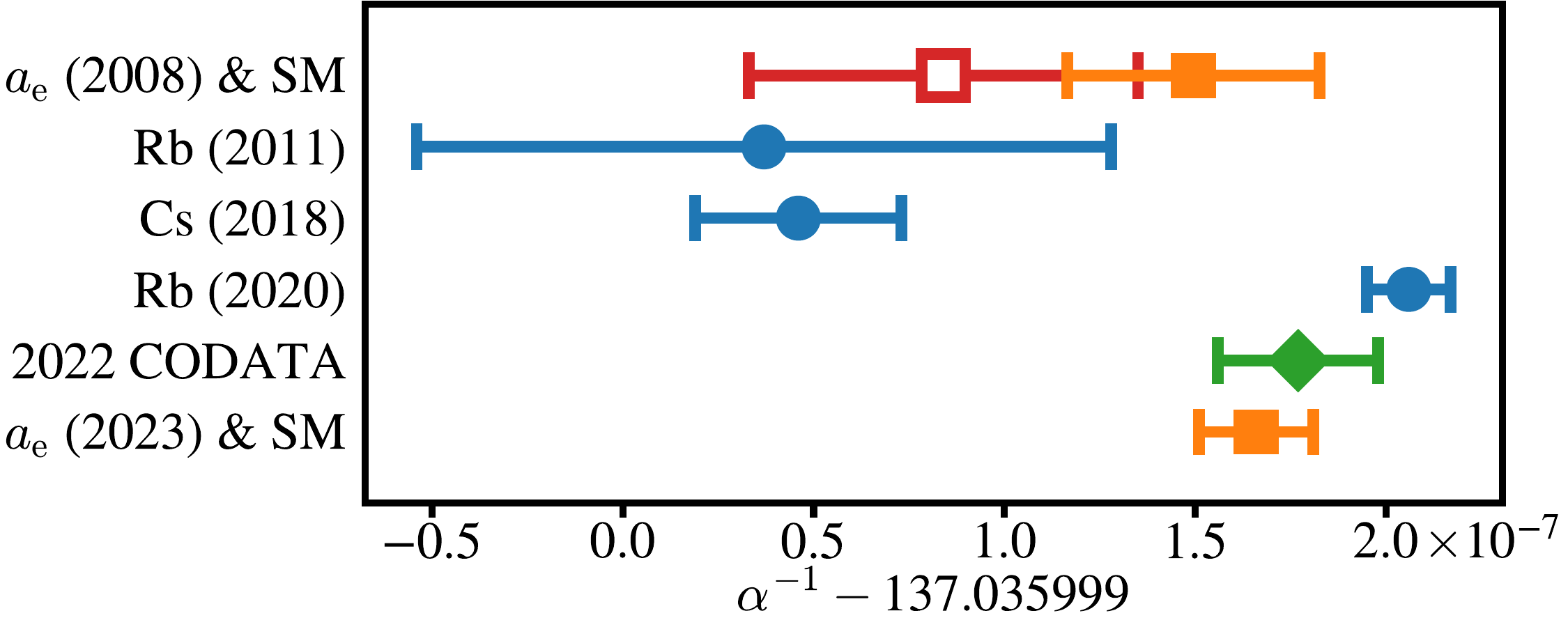}
    \caption{Recent high-precision determinations of the fine-structure constant. Blue circles represent values deduced using photon-recoil measurements from atom interferometry using Rb in 2011 \cite{Bouchendira2011} and 2020 \cite{Morel2020} and Cs in 2018 \cite{Parker2018}. Orange squares represent measurements of the anomalous magnetic moment of the electron ($a_{\mathrm{e}}$) from 2008 \cite{Hanneke2008} and 2023 \cite{Fan2023} combined with recent calculations from the Standard Model of particle physics \cite{Aoyama2019}. Using a different value for the 5-loop Feynman diagrams, from Ref. \cite{Volkov2019}, would shift both orange data points to the left by only $7 \times 10^{-9}$. The hollow red square represents the unadjusted determination from $a_{\mathrm{e}}$ in 2008 \cite{Hanneke2008}. The green diamond represents the 2022 CODATA recommended value \cite{CODATA2022} based on results published prior to 31 December 2022. Error bars represent $\pm 1\sigma$ for each datum.}
    \label{fig:alpha}
\end{figure}

Photon-recoil measurements from atom interferometry are used to determine the fine-structure constant via
\begin{equation}
    \alpha^2 = \frac{2 R_{\infty}}{c} \frac{A_r(X)}{A_r(\mathrm{e})} \frac{h}{m(X)},
\end{equation}
where $R_{\infty}$ is the Rydberg constant, $c$ is the speed of light, $A_r(X)$ is the relative mass of the atomic species $X$, $A_r(\mathrm{e})$ is the relative mass of the electron, $h$ is Planck's constant, and $m(X)$ is the absolute mass of $X$. The fine-structure constant is a free parameter of the Standard Model proportional to the square of the QED coupling constant and, thus, its determination is important for precision tests of fundamental physics. Recent high-precision determinations of $\alpha$ are presented in Fig. \ref{fig:alpha}. Determinations of $\alpha$ relying on atom interferometry are some of the most precise; Rb \cite{Morel2020} and Cs \cite{Parker2018} are currently the only two atoms used, but a discrepancy of more than 5\,$\sigma$ exists between the resulting values. This discrepancy limits the precision with which the electron magnetic moment can be compared with its Standard Model prediction \cite{Fan2023}. 

Its levity means that the electron is a factor of $(m_{\mu}/m_e)^2 \sim 4\times 10^4$ less sensitive to any lepton-flavour-universal beyond-Standard-Model interaction than the muon \cite{Giudice2012}. A possible tension between the current experimental and theoretical values of the anomalous magnetic moment of the muon exists at the level of $2.5\times 10^{-9}$ (see, e.g., \cite{Workman2022}); the electron would be sensitive to beyond-Standard-Model explanations of this tension if the measurement precision of its anomalous moment were improved by a factor of 2.2 \cite{Fan2023}, but exposing this would require the Standard Model prediction at a precision more than an order of magnitude better than offered by the current discrepancy in $\alpha$. Therefore, it is very pertinent to carry out additional high-precision determinations of the fine-structure constant; this could be achieved through additional atom interferometry experiments with, e.g., Sr, Yb (this work or \cite{PlotkinSwing2018,McAlpine2020}), or Li \cite{Rui2023}, and/or by other means, e.g. $g$ factor measurements of light ions \cite{Yerokhin2016} or He fine-structure measurements \cite{Heydarizadmotlagh2023}.

The uncertainty budget for the fine-structure constant (assuming negligible correlation between the experimental input values) comes down to
\begin{align}
    \left(\frac{\Delta\alpha}{\alpha}\right)^2 &\approx \frac{1}{4} \left[\left(\frac{\Delta R_{\infty}}{R_{\infty}}\right)^2 + \left(\frac{\Delta A_r(\mathrm{e})}{A_r(\mathrm{e})}\right)^2 \right.\nonumber\\
    &\quad\quad\left. + \left(\frac{\Delta A_r(X)}{A_r(X)}\right)^2 + \left(\frac{\Delta m(X)}{m(X)}\right)^2\right].
\end{align}

The current best relative uncertainties for the quantities other than the absolute mass of strontium are $\Delta R_{\infty}/R_{\infty} = 1.1\times10^{-12}$ \cite{CODATA2022}, $\Delta A_r(\mathrm{e})/A_r(\mathrm{e}) = 1.8\times 10^{-11}$ \cite{CODATA2022}, $\Delta A_r(\mathrm{Sr})/A_r(\mathrm{Sr}) = 7.0\times 10^{-11}$ \cite{Rana2012,Wang2021}. A measurement of the absolute mass of strontium at the level of precision of $\Delta m(\mathrm{Sr})/m(\mathrm{Sr}) \sim 1\times 10^{-11}$ would thus yield a value for the fine-structure constant at the level of precision of $\Delta\alpha/\alpha = 3.7\times 10^{-11}$, limited by the precision of the relative mass of strontium \cite{Rana2012,Wang2021}. This would represent an improvement in absolute mass measurement precision by an order of magnitude compared to Ref. \cite{Morel2020}, and would lead to an increase in the precision of the fine-structure constant by a factor of 2. Improvements in the precision of the relative masses, $A_r(\mathrm{e})$ and $A_r(\mathrm{Sr})$, to the level of $1\times 10^{-11}$ or below --- perhaps by using Penning traps operated at lower temperatures than previously \cite{Fan2024} 
--- would then lead to a precision for $\alpha$ an order of magnitude better than the current record.

For Yb, the relative uncertainty in the relative mass is $\Delta A_r(\mathrm{Yb})/A_r(\mathrm{Yb}) = 8.2\times 10^{-11}$ \cite{Rana2012,Wang2021}. Similarly, a measurement of the absolute mass of ytterbium at the level of precision of $\Delta m(\mathrm{Yb})/m(\mathrm{Yb}) \sim 1\times 10^{-11}$ would yield a precision of $\Delta\alpha/\alpha = 4.2\times 10^{-11}$ and be limited by the precision of the relative mass of Yb \cite{Rana2012,Wang2021}.

Other neutral divalent atomic species (alkaline-earth like atoms) that are being explored for use as optical clocks include Hg and Cd. The natural lifetimes of the clock state in ${}^{199,201}\mathrm{Hg}$ are $<2\,\mathrm{s}$, which would result in significant contrast loss from spontaneous emission for the method we propose. Although the natural lifetimes of the clock state in $^{111,113}\mathrm{Cd}$ are comparable to those in Yb, and the recoil frequencies are larger than for Sr by about a factor of 3, the precision to which the relative masses are known is worse by about a factor of 30. Accordingly, the increase in recoil frequency sensitivity afforded by the shorter clock wavelength motivates higher precision determinations of the Cd isotope relative masses before a competitive determination of $\alpha$ can be made based on them. The techniques and issues in making precise mass comparisons, including possible improvements, are reviewed by Myers \cite{Myers2013}.

Assuming an experimental phase resolution of $1\,\mathrm{mrad}$ \footnote{This could be achieved, for example, by averaging 100 shots at a resolution of $10\,\mathrm{mrad/shot}$}, we find an instrument of height $L=3\,\mathrm{m}$ operating in the `X' configuration with a laser pulse separation of $\dt = 1\,\mathrm{ms}$ to be sufficient for an absolute mass measurement of Sr or Yb at the level of $1\times 10^{-11}$. The total number of LMT pulses required for this, fewer than 500, is feasible with existing techniques \cite{Wilkason2022,Wilkason2022thesis}. Interestingly, such a precise measurement of the absolute mass of strontium would be sufficient to resolve the rest-mass difference between the ground and excited states,
\begin{equation}
    \frac{m(\mathrm{Sr}\ 5s5p\,{}^3P_0)}{m(\mathrm{Sr}\ 5s^2\,{}^1S_0)} - 1 = \frac{\hbar\omega}{m(\mathrm{Sr}\ 5s^2\,{}^1S_0) c^2} \approx 2.2 \times 10^{-11}.
\end{equation}

Reliably making an atom interferometry measurement of such high precision requires careful control of systematic effects. The majority of the systematic effects in \cite{Morel2020,Parker2018} arise from the laser beam. Intermediate-scale atom interferometers (prototypes for larger-scale gravitational wave detectors) are being designed to very high specifications \cite{Badurina2020,Abe2021} that should result in laser pulses with systematics comparable (at worst) to the present generation of Rb \cite{Morel2020} and Cs \cite{Parker2018} experiments (but there is scope to improve them). Although the light shifts and diffraction phases arising from Raman/Bragg pulses and Bloch oscillations used for Rb and Cs can be avoided by using single-photon pulses in Sr or Yb atoms (or other neutral atomic species with suitably narrow clock transitions), other systematics may arise. Further information about these is likely to be gained from the ongoing development work on the single-photon method.

Whilst we have used an LMT pulse separation of $\dt = 1\,\mathrm{ms} = \dtau$ to demonstrate the optimisation of trajectories, we note that a transform-limited pulse of such duration would have a bandwidth of less than 1\,kHz. This would require quite accurate pulse detunings in order to address the atoms, and so shorter pulses are likely to be preferred for an experimental implementation of our proposed scheme. Having established that sufficient precision can be achieved in principle with this approach, a future stage will be more complete modelling of the interferometry sequence that includes the frequency detuning of the pulses, and their selectivity (addressing one arm, or more). The theoretical methods for implementing this are well known \cite{Wilkason2022,Rudolph2020,Lellouch2023}.

The terms of order $1/c$ in (\ref{eq:dPhi}) due to the finite light propagation time are presented in Table \ref{tab:1/c magnitudes} for the trajectories in Sec. \ref{sec:case study}. Many of them are found to contribute more than $1\,\mathrm{mrad}$ to the total differential phase and thus represent a new systematic that arises when conducting offset simultaneous conjugate atom interferometry with macroscopic spatial separations.

\begin{table}[t]
    \centering
    \caption{Magnitude (in radians) of the terms in (\ref{eq:dPhi}) for example optimal trajectories found in Sec. \ref{sec:case study}. $L=5\,\mathrm{m}$ for Sr is given as an example for the `X' configuration, and $L=10\,\mathrm{m}$ for the fountain configuration. The indices are as in Table \ref{tab:breakdown}. Many terms contribute above the targeted $1\,\mathrm{mrad}$ level.}
    \label{tab:1/c magnitudes}
    \begin{ruledtabular}
    \begin{tabular}{lD{0}{0}{5.3}D{0}{0}{5.3}}
        Index & \multicolumn{1}{c}{`X' [L=5\,m]} & \multicolumn{1}{c}{Fountain [L=10\,m]} \\ \hline
        1 & 5\times 10^8 & 1\times 10^{10}\\
        2 & 4\times 10^7 & 1\times 10^9\\ 
        3 & 3\times 10^{-1} & 0\\
        4 & 2\times 10^{-1} & 4\times 10^{-3} \\
        5 & 6\times 10^{-1} & 2\times 10^{-1}\\
        6 & 9\times 10^{-2} & 9\times 10^{-1}\\
        7 & 9\times 10^{-3} & 2\times 10^{-1}\\
        8 & 2\times 10^{-2} & 1\times 10^{-2}\\
        9 & 3\times 10^{-2} & 2\times 10^{-1}\\ 
        10 & 6\times 10^{-3} & 1\times 10^{-1}\\ 
        11 & 0 & 5\times 10^{-6}\\
        12 & 0 & 2\times 10^{-5}\\
        13 & 0 & 5\times 10^{-7}\\
        14 & 0 & 0\\
        15 & 0 & 0\\
        16 & 2\times 10^{-7} & 5\times 10^{-7}
    \end{tabular}
    \end{ruledtabular}
\end{table} 

An even value of $N$ effectively results in the first $N+1$ pulses in the scheme forming a finite-time LMT beamsplitter in which the influenced arm is left in the excited state (as for a standard \RB\ interferometer). Using an odd value for $N$ does the same but leaves the influenced arm in the ground state. The parity of $M$ dictates whether or not the internal states are the same after the deflecting pulses as before. A modification to our scheme --- taking $M$ to be even for the diminished interferometer but odd for the augmented interferometer as well as sending pulses from the third \RB\ pulse onwards in both directions --- would allow its use with less long-lived atomic states whilst still benefiting from the long interaction times afforded by intermediate-scale atom interferometry instruments.

\section{\label{sec:conclusion}Conclusions}
Atom interferometry in intermediate-scale instruments using the long-lived excited states of single-photon clock transitions presents an opportunity for photon-recoil measurements with improved sensitivity. The technique of offset simultaneous conjugate \RB\ atom interferometry may be performed using ultracold atom clouds from separate sources. The ability to independently control the launch from each source will allow the trajectories of the atoms to be tailored in order to mitigate the gravity gradient phase while optimising the use of the interferometry region.

We propose a scheme utilising \pp s for LMT between the first and final pairs of \RB\ pulses to increase the photon-recoil energy difference and between the middle pair of \RB\ pulses to deflect the interferometers, the LMT order in these two parts of the interferometer being $N$ and $M$, respectively. The differential phase shift for this scheme is derived, including terms arising from the finite propagation time of the laser pulses between the interferometers, as well as an expression for the phase shift arising from the gravity gradient.

We find the optimal trajectories for the `X' configuration in an instrument of length $L=3\,\mathrm{m}$ for both Sr and Yb. Assuming an experimental resolution of $1\,\mathrm{mrad}$ can be achieved, then this optimised design of interferometer gives a recoil-frequency measurement that can improve the precision of the experimental fine-structure constant determination by a factor of 2 as compared to the current best precision \cite{Morel2020}. Such a measurement would also constitute the highest-precision absolute mass measurement of an atom, with a sufficient precision to resolve the mass difference between the strontium ground and excited states. Further gains in the precision of $\alpha$ would then be possible subject to improved relative mass measurements of the electron and atomic isotope, and will be necessary for future tests of the Standard Model by the electron magnetic moment.

An additional order of magnitude gain in absolute mass measurement precision would be possible in an instrument of size $L=10\,\mathrm{m}$, with further improvements for a fountain configuration. Further gains in sensitivity for any instrument size would come with shorter LMT pulses at sufficiently high Rabi frequencies.

While this design study is not based on any particular instrument or experiment, we note that there are 10-metre-scale atom interferometry experiments under development with Sr or Yb, including AION-10 at Oxford \cite{Badurina2020}, the Sr prototype tower at Stanford \cite{Abe2021}, and the VLBAI-Teststand in Hannover \cite{Hartwig2015}.

Finally, we note that the contributions to the phase shift arising from the propagation delay of the laser pulses between the interferometers can enter above the level of $1\,\mathrm{mrad}$ for the trajectories we have identified. These contributions are likely to become larger for even longer baselines, as proposed for atom interferometric gravitational-wave observatories such as AION \cite{Badurina2020}, MAGIS \cite{Abe2021}, MIGA \cite{Canuel2018}, ELGAR \cite{Canuel2020}, and ZAIGA \cite{Zhan2019}. Therefore, the detailed study of these effects considered in this paper by measurements on intermediate-scale prototypes is a crucial step towards very long baseline atom interferometry.

\begin{acknowledgments}
Thanks to Christopher McCabe for proof-reading an earlier version of this pre-print. This work was supported by UKRI through its Quantum Technology for Fundamental Physics programme, via the ST/T006633/1 grant from STFC in the framework of the AION Consortium. J.S. acknowledges support from the Rhodes Trust.
\end{acknowledgments}

\appendix
\renewcommand{\thesubsection}{\Alph{section}.\arabic{subsection}}
\renewcommand{\thesubsubsection}{\Alph{section}.\arabic{subsection}.\roman{subsubsection}}

\section{\label{app:maths} APPROXIMATIONS FOR CALCULATING TIME DIFFERENCES OF LIGHT BETWEEN TRAJECTORIES}
The derivations in the following appendices rely on some textbook mathematical results that are outlined below for convenient reference.

\subsection{\label{app:maths.quadratic} TAYLOR EXPANSION OF QUADRATIC FORMULA}
The quadratic formula, the general solution to the quadratic equation $0 = a x^2 + b x + c$, may be re-arranged as
\begin{equation}
    x = \frac{-b}{2 a} \left[1 \pm \sqrt{1 - \frac{4 a c}{b^2}} \right].
\end{equation}
For the case $b^2 \gg 4 a c$, the square-root may be approximated by its Taylor expansion
\begin{equation}
    x \approx \frac{-b}{2 a} \left[1 \pm \left(1 - \frac{2 a c}{b^2} - \frac{2 a^2 c^2}{b^4} \right) \right].
\end{equation}
Taking the negative sign (small $x$), we find
\begin{equation}
    x = \frac{-c}{b} + O\left(\frac{a c^2}{b^3}\right) \label{eq:quadratic_approximation}.
\end{equation}
Thus, to second order in $1/b$, the solution to the quadratic equation (for $b \gg a,c$) reduces to the solution to the linear equation $0 = b x + c$.

\subsection{\label{app:maths.geometric} GEOMETRIC SERIES}
The geometric series within the radius of convergence, $|r| < 1$, is given by
\begin{equation}
    \sum_{n=0}^{\infty} r^n = \frac{1}{1-r}.
\end{equation}
This result may be used to approximate the reciprocal of $a - x$ for the case $|a| \gg x$:
\begin{align}
    \frac{1}{a - x} &= \frac{1}{a} \left(\frac{1}{1-\frac{x}{a}}\right) \nonumber\\
    &\approx \frac{1}{a} \left(1 + \frac{x}{a} + \left(\frac{x}{a}\right)^2 \right) \label{eq:geometric_approximation}.
\end{align}

\subsection{\label{app:maths.Taylor} TAYLOR EXPANSION FOR ROOTS OF ANALYTIC FUNCTIONS}
Finding the time differences in (\ref{eq:z_beta}), (\ref{eq:z_gamma}), and (\ref{eq:z_delta}) compared to (\ref{eq:z_alpha})\ is equivalent to solving
\begin{align}
    0 = f(x_0 + \varepsilon) - g(x_0) \pm c \varepsilon \label{eq:for_Taylor_solution}
\end{align}
for $\varepsilon$, with functions $f$ and $g$ analytic at $x_0$, $\varepsilon \ll 1$, and $|f'(x_0)| \ll c$. An approximate solution to (\ref{eq:for_Taylor_solution}) may be found by taking the first-order Taylor expansion of $f$ about $x_0$, resulting in
\begin{align}
    0 &\approx f(x_0) + f'(x_0)\varepsilon - g(x_0) \pm c \varepsilon \nonumber\\
    \Rightarrow \varepsilon &\approx \frac{g(x_0) - f(x_0)}{\pm c + f'(x_0)}.
\end{align}
A bound on the error of the approximation can be found by taking the second-order Taylor expansion of $f$ about $x_0$ and using (\ref{eq:quadratic_approximation}):
\begin{align}
    0 &\approx \frac{f''(x_0)}{2} \varepsilon^2 + (f'(x_0) \pm c) \varepsilon + f(x_0) - g(x_0) \nonumber\\
    \Rightarrow \varepsilon &\approx \frac{g(x_0) - f(x_0)}{\pm c + f'(x_0)} + O\left(\frac{f''(x_0) (f(x_0) - g(x_0))^2}{c^3}\right).
\end{align}

A more useful approximation is found by applying the result of (\ref{eq:geometric_approximation}) to this expression. We find, to second-order in $1/c$,
\begin{align}
    \varepsilon &\approx (g(x_0) - f(x_0))\frac{1}{\pm c + f'(x_0)} \nonumber\\
    &\approx \frac{f(x_0) - g(x_0)}{c} \left(\mp 1 + \frac{f'(x_0)}{c}\right). \label{eq:Taylor_solution}
\end{align}

\subsection{\label{app:maths.result} APPLICATION TO TIME DIFFERENCES}
For our problem, we wish to solve for the intersection of the worldline of the photon with the worldline of the trajectory $z_j(t)$ ($j\in\{\beta,\gamma,\delta\}$) given coincidence of the photon with the trajectory $z_{\alpha}(t)$ at some time $t_i$. Ignoring general relativistic effects on the worldline of the photon, its trajectory is defined by
\begin{align}
    z_i(t) = z_{\alpha}(t_i) + \chi_i c (t - t_i).
\end{align}
The equation for the atom-photon intersection $z_j(t_i + \Delta t_i) = z_i(t_i + \Delta t_i)$ can be solved for $\Delta t_i$ by rearranging it to take the same form as (\ref{eq:for_Taylor_solution}):
\begin{align}
    0 &= z_j(t_i + \Delta t_i) - z_{\alpha}(t_i) - \chi_i c \Delta t_i. \label{eq:dt_definition}
\end{align}

\subsubsection{\label{app:maths.result.RB} \MakeUppercase{\RB}\ PULSES}
For the \RB\ pulses, $\Delta t_1$ is calculated for $z_{\gamma}(t)$ and $\Delta t_2$, $\Delta t_3$, and $\Delta t_4$ are calculated for $z_{\delta}(t)$. We find (to second-order in $1/c$)
\begin{align}
    0 &= z_{\gamma}(t_1 + \Delta t_1) - z_{\alpha}(t_1) - c \Delta t_1 \nonumber\\
    &= \widetilde{z_0}(t_1 + \Delta t_1) - z_0(t_1) - c \Delta t_1 \nonumber\\
    \Rightarrow \Delta t_1 &\approx \frac{1}{c} \left(1 + \frac{u + \Delta u - g t_1}{c}\right) \left(\Delta h + \Delta u t_1\right) \label{eq:dt1}
\end{align}
\begin{widetext}
\begin{align}
    0 &= z_{\delta}(t_2 + \Delta t_2) - z_{\alpha}(t_2) - c \Delta t_2 \nonumber\\
    &= \widetilde{z_0}(t_2 + \Delta t_2) - z_0(t_2) - v_r \left((N+1) T - 2 \sum_{n=1}^N t^{(n)}\right) - c \Delta t_2 \nonumber\\
    \Rightarrow \Delta t_2 &\approx \frac{1}{c} \left(1 + \frac{u + \Delta u - g t_2}{c}\right) \left(\Delta h + \Delta u t_2 - v_r \left((N+1) T - 2 \sum_{n=1}^N t^{(n)}\right)\right) \label{eq:dt2}
\end{align}
\end{widetext}
\onecolumngrid
\begin{align}
    0 &= z_{\delta}(t_3 + \Delta t_3) - z_{\alpha}(t_3) + (-1)^M c \Delta t_3 \nonumber\\
    &= \widetilde{z_0}(t_3 + \Delta t_3) + (M+1) v_r \Delta t_3 - z_0(t_3) + v_r \left((2M+1) (t_3 - t_2) - (N+1) T - \Delta t_2 - \sum_{m=1}^M \left(2 \tau^{(m)} + \delta\tau_{\delta}^{(m)}\right) \right.\nonumber\\
    &\quad\left. + 2 \sum_{n=1}^N t^{(n)} \right) + (-1)^M c \Delta t_3 \nonumber\\
    \Rightarrow \Delta t_3 &\approx \frac{1}{c} \left((-1)^{M+1} + \frac{u + \Delta u - g t_3 + (M+1) v_r}{c} \right) \Bigg(\Delta h + \Delta u t_3 + v_r \Bigg((2M+1) (t_3 - t_2) - (N+1) T - \Delta t_2 \nonumber\\
    &\quad - \sum_{m=1}^M \left(2 \tau^{(m)} + \delta\tau_{\delta}^{(m)}\right) + 2 \sum_{n=1}^N t^{(n)}\Bigg)\Bigg) \label{eq:dt3}
\end{align}
\begin{align}
    0 &= z_{\delta}(t_4 + \Delta t_4) - z_{\alpha}(t_4) + (-1)^M c \Delta t_4 \nonumber\\
    &= \widetilde{z_0}(t_4 + \Delta t_4) + (M+2) v_r \Delta t_4 - z_0(t_4) + v_r \Bigg((2M+1) (t_3 - t_2) + (N + 2 M + 2) T - \Delta t_3 - \Delta t_2 \nonumber\\
    &\quad - \sum_{m=1}^M \left(2 \tau^{(m)} + \delta\tau_{\delta}^{(m)}\right) - \sum_{n=1}^N \left(2 t^{(n)} - \Delta t_4^{(n)} + \Delta t_3^{(n)}\right) \Bigg) + (-1)^M c \Delta t_4 \nonumber\\
    \Rightarrow \Delta t_4 &\approx \frac{1}{c} \left((-1)^{M+1} + \frac{u + \Delta u - g t_4 + (M+2) v_r}{c} \right) \Bigg(\Delta h + \Delta u t_4 + v_r \Bigg((2M+1) (t_3 - t_2) + (N + 2 M + 2) T - \Delta t_3 \nonumber\\
    &\quad - \Delta t_2 - \sum_{m=1}^M \left(2 \tau^{(m)} + \delta\tau_{\delta}^{(m)}\right) - \sum_{n=1}^N \left(2 t^{(n)} - \Delta t_4^{(n)} + \Delta t_3^{(n)}\right) \Bigg)\Bigg) \label{eq:dt4}
\end{align}

\subsubsection{\label{app:maths.result.ARED} ADDITIONAL RECOIL-ENERGY-DIFFERENCE PULSES}
For the additional recoil-energy-difference pulses, $\Delta t_1^{(n)}$ and $\Delta t_2^{(n)}$ are calculated for $z_{\gamma}(t)$, and $\Delta t_3^{(n)}$ and $\Delta t_4^{(n)}$ are calculated for $z_{\delta}(t)$.

We find (to second-order in $1/c$)
\begin{align}
    0 &= z_{\gamma}\left(t_1 + t^{(n)} + \Delta t_1^{(n)}\right) - z_{\alpha}\left(t_1 + t^{(n)}\right) - (-1)^n c \Delta t_1^{(n)} \nonumber\\
    &= \widetilde{z_0}\left(t_1 + t^{(n)} + \Delta t_1^{(n)}\right) + n v_r \Delta t_1^{(n)} - z_0\left(t_1 + t^{(n)}\right) - v_r \left(\Delta t_1 + \sum_{p=1}^{n-1} \Delta t_1^{(p)} \right) - (-1)^n c \Delta t_1^{(n)} \nonumber\\
    \Rightarrow \Delta t_1^{(n)} &\approx \frac{1}{c} \left((-1)^n + \frac{u + \Delta u - g (t_1 + t^{(n)}) + n v_r}{c} \right) \left(\Delta h + \Delta u (t_1 + t^{(n)}) - v_r \left(\Delta t_1 + \sum_{p=1}^{n-1} \Delta t_1^{(p)} \right)\right) \label{eq:dt1n}
\end{align}
\begin{align}
    0 &= z_{\gamma}\left(t_2 - t^{(n)} + \Delta t_2^{(n)}\right) - z_{\alpha}\left(t_2 - t^{(n)}\right) - (-1)^n c \Delta t_2^{(n)} \nonumber\\
    &= \widetilde{z_0}\left(t_2 - t^{(n)} + \Delta t_2^{(n)}\right) + (n+1) v_r \Delta t_2^{(n)} - z_0\left(t_2 - t^{(n)}\right) - v_r \left(\Delta t_1 + \sum_{p=1}^N \Delta t_1^{(p)} - \sum_{p=n+1}^N \Delta t_2^{(p)} \right) \nonumber\\
    &\quad - (-1)^n c \Delta t_2^{(n)} \nonumber\\
    \Rightarrow \Delta t_2^{(n)} &\approx \frac{1}{c} \left((-1)^n + \frac{u + \Delta u - g (t_2 - t^{(n)}) + (n+1) v_r}{c} \right) \left(\Delta h + \Delta u (t_2 - t^{(n)}) - v_r \left(\Delta t_1 + \sum_{p=1}^N \Delta t_1^{(p)} \right.\right.\nonumber\\
    &\quad\left.\left. - \sum_{p=n+1}^N \Delta t_2^{(p)} \right)\right) \label{eq:dt2n}
\end{align}
\begin{align}
    0 &= z_{\delta}\left(t_3 + t^{(n)} + \Delta t_3^{(n)}\right) - z_{\alpha}\left(t_3 + t^{(n)}\right) + (-1)^{M+n} c \Delta t_3^{(n)} \nonumber\\
    &= \widetilde{z_0}\left(t_3 + t^{(n)} + \Delta t_3^{(n)}\right) + (M + n + 1) v_r \Delta t_3^{(n)} - z_0\left(t_3 + t^{(n)}\right) + v_r \Bigg((2 M + 1) (t_3 - t_2) - (N+1) T \nonumber\\
    &\quad + (2 M + 2 n + 1) t^{(n)} - \Delta t_3 - \Delta t_2 - \sum_{p=1}^{n-1} \Delta t_3^{(p)} - \sum_{m=1}^M \left(2 \tau^{(m)} + \delta\tau_{\delta}^{(m)}\right) + 2 \sum_{p=n}^N t^{(p)} \Bigg) + (-1)^{M+n} c \Delta t_3^{(n)} \nonumber\\
    \Rightarrow \Delta t_3^{(n)} &\approx \frac{1}{c} \left((-1)^{M+n+1} + \frac{u + \Delta u - g (t_3 + t^{(n)}) + (M + n + 1) v_r}{c} \right) \Bigg(\Delta h + \Delta u (t_3 + t^{(n)}) + v_r \Bigg((2 M + 1) (t_3 - t_2) \nonumber\\
    &\quad + (2 M + 2 n + 3) t^{(n)} - (N+1) T - \Delta t_3 - \Delta t_2 - \sum_{p=1}^{n-1} \Delta t_3^{(p)} - \sum_{m=1}^M \left(2 \tau^{(m)} + \delta\tau_{\delta}^{(m)}\right) + 2 \sum_{p=n+1}^N t^{(p)} \Bigg)\Bigg) \label{eq:dt3n}
\end{align}
\begin{align}
    0 &= z_{\delta}\left(t_4 - t^{(n)} + \Delta t_4^{(n)}\right) - z_{\alpha}\left(t_4 - t^{(n)}\right) + (-1)^{M+n} c \Delta t_4^{(n)} \nonumber\\
    &= \widetilde{z_0}\left(t_4 - t^{(n)} + \Delta t_4^{(n)}\right) + (M + n + 2) v_r \Delta t_4^{(n)} - z_0\left(t_4 - t^{(n)}\right) + v_r \Bigg((N + 2 M + 2) T - (2M + 2n + 3) t^{(n)} - \Delta t_3 \nonumber\\
    &\quad + (2 M + 1) (t_3 - t_2) - \Delta t_2 - \sum_{p=1}^N \Delta t_3^{(p)} - \sum_{p=n+1}^N \left(2 t^{(p)} - \Delta t_4^{(p)}\right) - \sum_{m=1}^M \left(2 \tau^{(m)} + \delta\tau_{\delta}^{(m)}\right) \Bigg) + (-1)^{M+n} c \Delta t_4^{(n)} \nonumber
\end{align}
\begin{align}
    \Rightarrow \Delta t_4^{(n)} &\approx \frac{1}{c} \left((-1)^{M+n+1} + \frac{u + \Delta u - g (t_4 - t^{(n)}) + (M + n + 2) v_r}{c} \right) \Bigg(\Delta h + \Delta u (t_4 - t^{(n)}) + v_r \Bigg((N + 2 M + 2) T \nonumber\\
    &\quad + (2 M + 1) (t_3 - t_2) - (2M + 2n + 3) t^{(n)} - \Delta t_3 - \Delta t_2 - \sum_{p=n+1}^N \left(2 t^{(p)} - \Delta t_4^{(p)}\right) - \sum_{m=1}^M \left(2 \tau^{(m)} + \delta\tau_{\delta}^{(m)}\right) \nonumber\\
    &\quad - \sum_{p=1}^N \Delta t_3^{(p)} \Bigg)\Bigg) \label{eq:dt4n}
\end{align}

\subsubsection{\label{app:maths.result.deflecting} DEFLECTING PULSES}
For the deflecting pulses, $\delta\tau_{\beta}^{(m)}$ is calculated for $z_{\beta}(t)$, $\delta\tau_{\gamma}^{(m)}$ for $z_{\gamma}(t)$, and $\delta\tau_{\delta}^{(m)}$ for $z_{\delta}(t)$.

We find (to second-order in $1/c$)
\begin{align}
    0 &= z_{\beta}\left(t_2 + \tau^{(m)} + \delta\tau_{\beta}^{(m)}\right) - z_{\alpha}\left(t_2 + \tau^{(m)}\right) - (-1)^m c \delta\tau_{\beta}^{(m)} \nonumber\\
    &= z_0\left(t_2 + \tau^{(m)} + \delta\tau_{\beta}^{(m)}\right) - (m-1) v_r \delta\tau_{\beta}^{(m)} - z_0\left(t_2 + \tau^{(m)}\right) - v_r \left((N+1) T - \sum_{p=1}^{m-1} \delta\tau_{\beta}^{(p)} - 2 \sum_{n=1}^{N} t^{(n)} \right) \nonumber\\
    &\quad - (-1)^m c \delta\tau_{\beta}^{(m)} \nonumber\\
    \Rightarrow \delta\tau_{\beta}^{(m)} &\approx \frac{v_r}{c} \left((-1)^m + \frac{u - g (t_2 + \tau^{(m)}) - (m-1) v_r}{c} \right) \left( - (N+1) T + \sum_{p=1}^{m-1} \delta\tau_{\beta}^{(p)} + 2 \sum_{n=1}^{N} t^{(n)}\right) \label{eq:dtau_beta^m}
\end{align}
\begin{align}
    0 &= z_{\gamma}\left(t_2 + \tau^{(m)} + \delta\tau_{\gamma}^{(m)}\right) - z_{\alpha}\left(t_2 + \tau^{(m)}\right) - (-1)^m c \delta\tau_{\gamma}^{(m)} \nonumber\\
    &= \widetilde{z_0}\left(t_2 + \tau^{(m)} + \delta\tau_{\gamma}^{(m)}\right) + m v_r \delta\tau_{\gamma}^{(m)} - z_0\left(t_2 + \tau^{(m)}\right) + v_r \left((2 m - 1) \tau^{(m)} - \Delta t_1 + \sum_{n=1}^N \left(\Delta t_2^{(n)} - \Delta t_1^{(n)}\right) \right.\nonumber\\
    &\quad\left. - \sum_{p=1}^{m-1} \left(2 \tau^{(p)} + \delta\tau_{\gamma}^{(p)}\right) \right) - (-1)^m c \delta\tau_{\gamma}^{(m)} \nonumber\\
    \Rightarrow \delta\tau_{\gamma}^{(m)} &\approx \frac{1}{c} \left((-1)^m + \frac{u + \Delta u - g (t_2 + \tau^{(m)}) + m v_r}{c} \right) \Bigg(\Delta h + \Delta u (t_2 + \tau^{(m)}) + v_r \Bigg((2 m - 1) \tau^{(m)} - \Delta t_1 \nonumber\\
    &\quad + \sum_{n=1}^N \left(\Delta t_2^{(n)} - \Delta t_1^{(n)}\right) - \sum_{p=1}^{m-1} \left(2 \tau^{(p)} + \delta\tau_{\gamma}^{(p)}\right) \Bigg)\Bigg) \label{eq:dtau_gamma^m}
\end{align}
\begin{align}
    0 &= z_{\delta}\left(t_2 + \tau^{(m)} + \delta\tau_{\delta}^{(m)}\right) - z_{\alpha}\left(t_2 + \tau^{(m)}\right) - (-1)^m c \delta\tau_{\delta}^{(m)} \nonumber\\
    &= \widetilde{z_0}\left(t_2 + \tau^{(m)} + \delta\tau_{\delta}^{(m)}\right) + m v_r \delta\tau_{\delta}^{(m)} - z_0\left(t_2 + \tau^{(m)}\right) + v_r \left((2 m - 1) \tau^{(m)} - \Delta t_2 - \sum_{p=1}^{m-1} \left(2 \tau^{(p)} + \delta\tau_{\delta}^{(p)}\right) \right.\nonumber\\
    &\quad\left. - (N+1) T + 2 \sum_{n=1}^N t^{(n)} \right) - (-1)^m c \delta\tau_{\delta}^{(m)} \nonumber\\
    \Rightarrow \delta\tau_{\delta}^{(m)} &\approx \frac{1}{c} \left((-1)^m + \frac{u + \Delta u - g (t_2 + \tau^{(m)}) + m v_r}{c} \right) \Bigg(\Delta h + \Delta u (t_2 + \tau^{(m)}) + v_r \Bigg((2 m - 1) \tau^{(m)} - \Delta t_2 \nonumber\\
    &\quad - \sum_{p=1}^{m-1} \left(2 \tau^{(p)} + \delta\tau_{\delta}^{(p)}\right) - (N+1) T + 2 \sum_{n=1}^N t^{(n)} \Bigg)\Bigg) \label{eq:dtau_delta^m}
\end{align}

\begin{table*}[b]
    \centering
    \caption{Propagation delays to first-order in $1/c$ for $t^{(n)} = n \dt$ and $\tau^{(m)} = t_d - t_2 + (m-1) \dt$. We define $\mathfrak{t}(t) = t - t_2 + M (2 (t - t_d) - (M-1)\dt)$ and $\mathfrak{N} = (N+1)(T - N \dt)$ to simplify the expressions.}
    \label{tab:1/c delays}
    \begin{ruledtabular}
    \begin{tabular}{ll}
        Symbol & Expression \\\hline
        $c \Delta t_1$ & $\Delta h + \Delta u t_1$ \\
        $c \Delta t_2$ & $\Delta h + \Delta u t_2 - \mathfrak{N} v_r$ \\
        $c \Delta t_3$ & $(-1)^{M+1} \left(\Delta h + \Delta u t_3 + \left(\mathfrak{t}(t_3) - \mathfrak{N}\right) v_r \right)$ \\
        $c \Delta t_4$ & $(-1)^{M+1} \left(\Delta h + \Delta u t_4 + \left(\mathfrak{t}(t_4) + \mathfrak{N}\right) v_r \right)$ \\
        $c \Delta t_1^{(n)}$ & $(-1)^n \left(\Delta h + \Delta u (t_1 + n \dt)\right)$ \\
        $c \Delta t_2^{(n)}$ & $(-1)^n \left(\Delta h + \Delta u (t_2 - n \dt)\right)$ \\
        $c \Delta t_3^{(n)}$ & $(-1)^{M+n+1} \left(\Delta h + \Delta u (t_3 + n \dt) + \left(\mathfrak{t}(t_3) + n (2 M + n + 2) \dt - \mathfrak{N}\right) v_r \right)$ \\
        $c \Delta t_4^{(n)}$ & $(-1)^{M+n+1} \left(\Delta h + \Delta u (t_4 - n \dt) + \left(\mathfrak{t}(t_4) - n (2M + n + 2) \dt + \mathfrak{N} \right) v_r \right)$ \\
        $c \delta\tau_{\beta}^{(m)}$ & $(-1)^{m+1} \mathfrak{N} v_r$ \\
        $c \delta\tau_{\gamma}^{(m)}$ & $(-1)^m \left(\Delta h + \Delta u (t_d + (m-1) \dt) + \left(t_d - t_2 + (m+1) (m-1) \dt \right) v_r \right)$ \\
        $c \delta\tau_{\delta}^{(m)}$ & $(-1)^m \left(\Delta h + \Delta u (t_d + (m-1) \dt) + \left(t_d - t_2 + (m+1) (m-1) \dt - \mathfrak{N} \right) v_r \right)$
    \end{tabular}
    \end{ruledtabular}
\end{table*}

The time delays carrying a factor of $c$ require substitution of the full second-order expressions (\ref{eq:dt1}) -- (\ref{eq:dtau_delta^m}) except for the terms arising from time delays that, since they are of order $1/c$, introduce terms at second-order in $1/c$ in the final phase. All other time delay terms need only be substituted as the first-order expressions presented in Table \ref{tab:1/c delays}. It is useful to note that, to first-order, $\delta\tau_{\delta}^{(m)} = \delta\tau_{\gamma}^{(m)} + \delta\tau_{\beta}^{(m)}$.

\newpage
\section{\label{app:b}DERIVATION OF PHASE FOR THE PROPOSED ATOM INTERFEROMETER}
After substituting the values from Table \ref{tab:all_pulses} into (\ref{eq:DPhi}), noting that the terms with $\phi_i$ do not contribute, we find

\begin{align}
    \Delta\Phi &= k \Bigg(\overline{z}_a(t_1 + \Delta t_1) - \overline{z}_d(t_1) - \overline{z}_a(t_2 + \Delta t_2) + \overline{z}_d(t_2) - \overline{z}_a(t_3 + \Delta t_3) + \overline{z}_d(t_3) + \overline{z}_a(t_4 + \Delta t_4) - \overline{z}_d(t_4) \nonumber\\
    &\quad + \sum_{n=1}^N \bigg[\overline{z}_a\left(t_1 + t^{(n)} + \Delta t_1^{(n)}\right) - \overline{z}_d\left(t_1 + t^{(n)}\right) - \overline{z}_a\left(t_2 - t^{(n)} + \Delta t_2^{(n)}\right) + \overline{z}_d\left(t_2 - t^{(n)}\right) \nonumber\\
    &\quad\quad\quad\; - \overline{z}_a\left(t_3 + t^{(n)} + \Delta t_3^{(n)}\right) + \overline{z}_d\left(t_3 + t^{(n)}\right) + \overline{z}_a\left(t_4 - t^{(n)} + \Delta t_4^{(n)}\right) - \overline{z}_d\left(t_4 - t^{(n)}\right)\bigg] \nonumber\\
    &\quad + \sum_{m=1}^M \bigg[\overline{z}_a\left(t_2 + \tau^{(m)} + \delta\tau_{\gamma}^{(m)}\right) - \overline{z}_a\left(t_2 + \tau^{(m)} + \delta\tau_{\delta}^{(m)}\right) + \overline{z}_d\left(t_2 + \tau^{(m)}\right) - \overline{z}_d\left(t_2 + \tau^{(m)} + \delta\tau_{\beta}^{(m)}\right)\bigg]\Bigg) \label{eq:derivation_part_1}
\end{align}

Using (\ref{eq:midpoint_za}) and (\ref{eq:midpoint_zd}), (\ref{eq:derivation_part_1}) can be re-written in terms of light propagation delays (\ref{eq:dt_definition}) and the arm separations of the interferometers:

\begin{align}
    \Delta\Phi &= k c \left(\Delta t_1 - \Delta t_2 + (-1)^M \left(\Delta t_3 - \Delta t_4\right) + \sum_{n=1}^N (-1)^n \left(\Delta t_1^{(n)} - \Delta t_2^{(n)} + (-1)^M \left(\Delta t_3^{(n)} - \Delta t_4^{(n)}\right) \right) \right.\nonumber\\
    &\quad\left. + \sum_{m=1}^M (-1)^m\left(\delta\tau_{\gamma}^{(m)} - \delta\tau_{\delta}^{(m)} - \delta\tau_{\beta}^{(m)}\right)\right) + \frac{k}{2} \Bigg(\Delta z_d(t_1) - \Delta z_a(t_1 + \Delta t_1) - \left(\Delta z_d(t_2) + \Delta z_a(t_2 + \Delta t_2)\right) \nonumber\\
    &\quad\quad - \left(\Delta z_d(t_3) + \Delta z_a(t_3 + \Delta t_3)\right) + \Delta z_d(t_4) + \Delta z_a(t_4 + \Delta t_4) \nonumber\\
    &\quad + \sum_{n=1}^N \bigg[\Delta z_d\left(t_1 + t^{(n)}\right) - \Delta z_a\left(t_1 + t^{(n)} + \Delta t_1^{(n)}\right) - \left(\Delta z_d\left(t_2 - t^{(n)}\right) - \Delta z_a\left(t_2 - t^{(n)} + \Delta t_2^{(n)}\right)\right) \nonumber\\
    &\quad\quad\ \  - \left(\Delta z_d\left(t_3 + t^{(n)}\right) + \Delta z_a\left(t_3 + t^{(n)} + \Delta t_3 ^{(n)}\right)\right) + \Delta z_d\left(t_4 - t^{(n)}\right) + \Delta z_a\left(t_4 - t^{(n)} + \Delta t_4^{(n)}\right) \bigg] \nonumber\\
    &\quad - \sum_{m=1}^M \left[\Delta z_d\left(t_2 + \tau^{(m)}\right) + \Delta z_a \left(t_2 + \tau^{(m)} + \delta\tau_{\gamma}^{(m)}\right) + \Delta z_d\left(t_2 + \tau^{(m)} + \delta\tau_{\beta}^{(m)}\right) + \Delta z_a\left(t_2 + \tau^{(m)} + \delta\tau_{\delta}^{(m)}\right)\right]\Bigg) \label{eq:derivation_part_2}
\end{align}

The terms in (\ref{eq:derivation_part_2}) that carry a factor of $c$ are those that require the use of the second-order time delays in (\ref{eq:dt1}) -- (\ref{eq:dtau_delta^m}). To simplify this expression, we note that for laser pulses $i$ and $j$ the sum and difference of the impulse response functions accounting for propagation are given by

\begin{align}
    \zeta(t_i,t_j) - \zeta(t_i + \Delta t_1, t_j + \Delta t_j) &= v_r \theta(t_i - t_j) (t_i - t_j) - v_r \theta(t_i + \Delta t_i - t_j - \Delta t_j) (t_i + \Delta t_i - t_j - \Delta t_j) \nonumber\\
    &= v_r \theta(t_i - t_j) (\Delta t_j - \Delta t_i) \label{eq:zeta_difference}\\
    \zeta(t_i,t_j) + \zeta(t_i + \Delta t_1, t_j + \Delta t_j) &= v_r \theta(t_i - t_j) (t_i - t_j) + v_r \theta(t_i + \Delta t_i - t_j - \Delta t_j) (t_i + \Delta t_i - t_j - \Delta t_j) \nonumber\\
    &= v_r \theta(t_i - t_j) (2(t_i - t_j) + \Delta t_i - \Delta t_j), \label{eq:zeta_sum}
\end{align}
where $\theta(t)$ is the Heaviside step function. Hence, (\ref{eq:zeta_difference}) and (\ref{eq:zeta_sum}) can be used to express the sum or difference between arm separations at the times of the laser pulses:

\begin{align}
    &\Delta z_d(t_i) - \Delta z_a(t_i + \Delta t_i) \nonumber\\
    &\approx v_r \Bigg(\theta(t_i - t_1) (\Delta t_1 - \Delta t_i) - \theta(t_i - t_2) (\Delta t_2 - \Delta t_i) - \theta(t_i - t_3) (\Delta t_3 - \Delta t_i) \nonumber\\
    &\quad\quad + \sum_{n=1}^N \Big[\theta\left(t_i - t_1 - t^{(n)}\right) \left(\Delta t_1^{(n)} - \Delta t_i\right) - \theta\left(t_i - t_2 + t^{(n)}\right) \left(\Delta t_2^{(n)} - \Delta t_i\right) - \theta\left(t_i - t_3 - t^{(n)}\right) \left(\Delta t_3^{(n)} - \Delta t_i\right) \nonumber\\
    &\quad\quad\quad + \theta\left(t_i - t_4 + t^{(n)}\right) \left(\Delta t_4^{(n)} - \Delta t_i\right)\Big] - \sum_{m=1}^M \Big[\theta\left(t_i - t_2 - \tau^{(m)}\right) \left(2(t_i - t_2 - \tau^{(m)}) + \Delta t_i - \delta\tau_{\gamma}^{(m)}\right) \nonumber\\
    &\quad\quad\quad - \theta\left(t_i - t_2 - \tau^{(m)} - \delta\tau_{\beta}^{(m)}\right) \left(2(t_i - t_2 - \tau^{(m)} - \delta\tau_{\beta}^{(m)}) + \Delta t_i - \delta\tau_{\gamma}^{(m)}\right)\Big] \Bigg) \label{eq:dz_difference}\\
    &\Delta z_d(t_i) + \Delta z_a(t_i + \Delta t_i) \nonumber\\
    &\approx v_r \Bigg(\theta(t_i - t_1) (2(t_i - t_1) + \Delta t_i - \Delta t_1) - \theta(t_i - t_2) (2(t_i - t_2) + \Delta t_i - \Delta t_2) - \theta(t_i - t_3) (2(t_i - t_3) + \Delta t_i - \Delta t_3) \nonumber\\
    &\quad + \sum_{n=1}^N \Big[\theta\left(t_i - t_1 - t^{(n)}\right) \left(2(t_i - t_1 - t^{(n)}) + \Delta t_i - \Delta t_1^{(n)}\right) - \theta\left(t_i - t_2 + t^{(n)}\right) \left(2(t_i - t_2 + t^{(n)}) + \Delta t_i - \Delta t_2^{(n)}\right) \nonumber\\
    &\quad\quad - \theta\left(t_i - t_3 - t^{(n)}\right) \left(2(t_i - t_3 - t^{(n)}) + \Delta t_i - \Delta t_3^{(n)}\right) + \theta\left(t_i - t_4 + t^{(n)}\right) \left(2(t_i - t_4 + t^{(n)}) + \Delta t_i - \Delta t_4^{(n)}\right)\Big]\Bigg), \label{eq:dz_sum}
\end{align}
where we have used the result that $\delta\tau_{\delta}^{(m)} = \delta\tau_{\gamma}^{(m)} + \delta\tau_{\beta}^{(m)}$ to first-order in $1/c$. After substituting (\ref{eq:dz_difference}) and (\ref{eq:dz_sum}) into (\ref{eq:derivation_part_2}) and simplifying, we find

\begin{align}
    \Delta\Phi &= k c \left(\Delta t_1 - \Delta t_2 + (-1)^M \left(\Delta t_3 - \Delta t_4\right) + \sum_{n=1}^N (-1)^n \left(\Delta t_1^{(n)} - \Delta t_2^{(n)} + (-1)^M \left(\Delta t_3^{(n)} - \Delta t_4^{(n)}\right) \right) \right.\nonumber\\
    &\quad\left. + \sum_{m=1}^M (-1)^m\left(\delta\tau_{\gamma}^{(m)} - \delta\tau_{\delta}^{(m)} - \delta\tau_{\beta}^{(m)}\right)\right) + \omega_r \Bigg(- 2 (N+1) (N+2M+2) T + (2M+1) \left(\Delta t_1 - \Delta t_2\right) \nonumber\\
    &\quad + \Delta t_3 - \Delta t_4 + \sum_{n=1}^N \Bigg[4 (2M+n+2) t^{(n)} + (2M-n+1) \left(\Delta t_1^{(n)} - \Delta t_2^{(n)}\right) + (n+1) \left(\Delta t_3^{(n)} - \Delta t_4^{(n)}\right) \nonumber\\
    &\quad + \sum_{p=n}^N \left[4 t^{(p)} + \Delta t_2^{(p)} - \Delta t_1^{(p)} + \Delta t_3^{(p)} - \Delta t_4^{(p)}\right] \Bigg] \Bigg), \label{eq:derivation_part_4}
\end{align}
where $\omega_r = \hbar k^2 / (2m)$ is the atomic photon-recoil frequency. This is equation (\ref{eq:derivation_result_main_text}) in the main text. We use the LMT pulse timings defined by (\ref{eq:t^n}) and (\ref{eq:tau^m}), and the propagation delays defined by (\ref{eq:dt1}) -- (\ref{eq:dtau_delta^m}) and Table \ref{tab:1/c delays} to evaluate (\ref{eq:derivation_part_4}) using \textsc{mathematica}, with the case of even $N$ and $M$ verified by hand. We find

\begin{align}
    \Delta\Phi &= (N+1) (N + 2M + 2) k v_r (T - N \dt) - \frac{N (N+1) (N+2) k v_r \dt}{3} \nonumber\\
    &\quad + \frac{(1 + (-1)^M) k g \Delta h \mathcal{T}}{2 c} - \frac{k \Delta u \mathcal{T}}{2 c} \left[u + \Delta u - g (t_1 + t_2) + (-1)^M \left(u + \Delta u - g (t_3 + t_4) + (M+1) v_r\right)\right] \nonumber\\
    &\quad - \frac{(-1)^M k v_r \mathcal{T}}{2 c} \left[(2M+1) \left(u + \Delta u - g (t_4 + t_3 - t_d) + (M+1) v_r\right) - g \left(t_d - t_2 - M^2 \dtau\right)\right] \nonumber\\
    &\quad + \frac{k v_r \left(T - N \dt\right)}{2 c} \left[(N+1) \left(\Delta u + g (2 (t_d - t_2) - \dtau)\right) \right.\nonumber\\
    &\quad\quad\left. - (-1)^M \left((N+1) \left(\Delta u - g (t_4 + t_3 - 2 t_d + \dtau) + (2M+1) v_r\right) + 2 N M g \dtau \right) \right]\nonumber\\
    &\quad - \frac{(-1)^N k v_r \left(T - (2N+1)\dt\right)}{2 c} \left[(-1)^M (N+1) \left(2 u + 3 \Delta u - g (t_3 + t_4) + (2N+4M+5) v_r\right) \right.\nonumber\\
    &\quad\quad\left. + (N+1) \Delta u + (-1)^M (1 - (-1)^N) M g \dtau \right]\nonumber\\
    &\quad - \frac{k v_r \dt }{4 c} \left[(1 + (-1)^N) \left(\Delta u + (-1)^M \left(2 u + 3 \Delta u - g (t_3 + t_4) + (4M+3) v_r\right)\right) \right.\nonumber\\
    &\quad\quad\left. + 4 (-1)^{N+M} (N+1) v_r\right], \label{eq:dPhi_full}
\end{align}
where $\mathcal{T} = T + \dt + (-1)^N \left(T - (2N+1)\dt\right)$. This is the full expression for (\ref{eq:dPhi}) including the terms of order $1/c$. Its terms are indexed by their position and deconstructed in Table \ref{tab:breakdown}.

\begin{table*}[b]
    \centering
    \caption{Breakdown of (\ref{eq:dPhi_full}) into its component terms. Each term is ascribed an index according to its position in the expression. We define $\mathcal{T} = T + \dt + (-1)^N \left(T - (2N+1)\dt\right)$ for brevity. The dimensionless ratios are all either unity or the ratio of quantities with dimensions of speed.}
    \label{tab:breakdown}
    \begin{ruledtabular}
    \begin{tabular}{lllll}
        Index & Angular frequency & Time & Dimensionless ratio & Numerical coefficient \\ \hline
        1 & $k v_r$ & $T - N \dt$ & 1 & $(N+1) (N + 2 M + 2)$\\
        2 & $k v_r$ & $\dt$ & $1$ & $-N (N+1) (N + 2)/3$ \\ 
        3 & $k g \Delta h/c$ & $\mathcal{T}$ & 1 & $(1 + (-1)^M)/2$\\
        4 & $k \Delta u$ & $\mathcal{T}$ & $\left(u + \Delta u - g (t_1 + t_2)\right)/c$ & $-1/2$ \\
        5 & $k \Delta u$ & $\mathcal{T}$ & $\left(u + \Delta u - g (t_3 + t_4) + (M+1) v_r\right)/c$ & $(-1)^{M+1}/2$ \\
        6 & $k v_r$ & $\mathcal{T}$ & $\left(u + \Delta u - g (t_4 + t_3 - t_d) + (M+1) v_r\right)/c$ & $(-1)^{M+1} (2M+1)/2$\\
        7 & $k v_r$ & $\mathcal{T}$ & $g \left(t_d - t_2 - M^2 \dtau\right)/c$ & $(-1)^M/2$\\
        8 & $k v_r$ & $T - N \dt$ & $\left(\Delta u + g (2 (t_d - t_2) - \dtau)\right)/c$ & $(N+1)/2$\\
        9 & $k v_r$ & $T - N \dt$ & $\left(\Delta u - g (t_4 + t_3 - 2 t_d + \dtau) + (2M+1) v_r\right)/c$ & $(-1)^{M+1} (N+1)/2$\\
        10 & $k v_r$ & $T - N \dt$ & $g \dtau / c$ & $(-1)^{M+1} N M$\\
        11 & $k v_r$ & $T - (2N+1)\dt$ & $\left(2u + 3 \Delta u - g (t_3 + t_4) + (2N+4M+5) v_r\right)/c$ & $(-1)^{N+M+1} (N+1)/2$\\
        12 & $k v_r$ & $T - (2N+1)\dt$ & $\Delta u / c$ & $(-1)^{N+1} (N+1)/2$\\
        13 & $k v_r$ & $T - (2N+1)\dt$ & $g \dtau / c$ & $(-1)^{M+1} (1 - (-1)^N) M/2$\\
        14 & $k v_r$ & $\dt$ & $\Delta u / c$ & $-(1 + (-1)^N)/4$\\
        15 & $k v_r$ & $\dt$ & $\left(2 u + 3 \Delta u - g (t_3 + t_4) + (4M+3) v_r\right)/c$ & $(-1)^{M+1} (1 + (-1)^N)/4$\\
        16 & $k v_r$ & $\dt$ & $v_r/c$ & $(-1)^{N+M+1} (N+1)$ 
    \end{tabular}
    \end{ruledtabular}
\end{table*}

\section{\label{app:c}DERIVATION OF GRAVITY GRADIENT PHASE FOR OUR ATOM INTERFEROMETER}
After substituting equations (\ref{eq:dz}) and (\ref{eq:zp-zd}) into (\ref{eq:differential_gravity_gradient_phase_approx_1}) and using (\ref{eq:t^n}) and (\ref{eq:tau^m}), we partition the integral for $\Delta\Phi_{\mathrm{GG}}$ into time intervals between consecutive laser pulses to find
\begin{align}
    \Delta\Phi_{\mathrm{GG}} &\approx - k G_{zz} \left[\sum_{n=1}^N \left[\int_{t_1 + (n-1) \dt}^{t_1 + n \dt} \left(\Delta h + \Delta u t\right) \left(n (t - t_1) - \dt \sum_{p=1}^{n-1} p \right) dt\right] \right.\nonumber\\
    &\quad\quad\quad\left. + \int_{t_1 + N \dt}^{t_2 - N \dt} \left(\Delta h + \Delta u t\right) \left((N+1) (t - t_1) - \dt \sum_{n=1}^N n \right) dt \right.\nonumber\\
    &\quad\quad\quad\left. + \sum_{n=1}^N \left[\int_{t_2 - n \dt}^{t_2 - (n-1) \dt} \left(\Delta h + \Delta u t\right) \left((N+1) T + n (t - t_2) - \dt \left(\sum_{p=1}^N p + \sum_{p=n}^N p\right) \right) dt\right] \right.\nonumber\\
    &\quad\quad\quad\left. + \int_{t_2}^{t_d} \left(\Delta h + \Delta u t + v_r (t - t_2)\right) \left((N+1) T - 2 \dt \sum_{n=1}^N n \right) dt \right.\nonumber\\
    &\quad\quad\quad\left. + \sum_{m=1}^{M-1} \left[\int_{t_d + (m-1) \dtau}^{t_d + m \dtau} \left(\Delta h + \Delta u t + v_r \left(t - t_2 + 2 m (t - t_d) - 2 \dtau \sum_{p=1}^m (p-1)\right)\right) \right.\right.\nonumber\\
    &\quad\quad\quad\quad\quad\left.\left. \times \left((N+1) T - 2 \dt \sum_{n=1}^N n \right) dt\right] \right.\nonumber\\
    &\quad\quad\quad\left. + \int_{t_d + (M-1) \dtau}^{t_3} \left(\Delta h + \Delta u t + v_r \left(t - t_2 + 2 M (t - t_d) - 2 \dtau \sum_{m=1}^M (m-1)\right)\right) \right.\nonumber\\
    &\quad\quad\quad\quad\quad\left. \times \left((N+1) T - 2 \dt \sum_{n=1}^N n \right) dt \right.\nonumber\\
    &\quad\quad\quad\left. + \sum_{n=1}^N \left[\int_{t_3 + (n-1) \dt}^{t_3 + n \dt} \left(\Delta h + \Delta u t + v_r \left(t - t_2 + 2 M (t - t_d) + n (t - t_3) - \dt \sum_{p=1}^{n-1} p \right.\right.\right.\right.\nonumber\\
    &\quad\quad\quad\quad\quad\left.\left.\left.\left. - 2 \dtau \sum_{m=1}^M (m-1)\right)\right) \left((N+1) T - n (t - t_3) - \dt \left(\sum_{p=1}^N p + \sum_{p=n}^N p\right) \right) dt\right] \right.\nonumber\\
    &\quad\quad\quad\left. + \int_{t_3 + N \dt}^{t_4 - N \dt} \left(\Delta h + \Delta u t + v_r \left(t - t_2 + 2 M (t - t_d) + (N+1) (t - t_3) - \dt \sum_{n=1}^N n \right.\right.\right.\nonumber\\
    &\quad\quad\quad\quad\quad\left.\left.\left. - 2 \dtau \sum_{m=1}^M (m-1)\right)\right) \left((N+1) (T - (t - t_3)) - \dt \sum_{n=1}^N n \right) dt \right.\nonumber\\
    &\quad\quad\quad\left. + \sum_{n=1}^N \left[\int_{t_4 - n \dt}^{t_4 - (n-1) \dt} \left(\Delta h + \Delta u t + v_r \left(t - t_2 + 2 M (t - t_d)  + (N+1) T + n (t - t_4) \right.\right.\right.\right.\nonumber\\
    &\quad\quad\quad\quad\quad\left.\left.\left.\left. - \dt \left(\sum_{p=1}^N p + \sum_{p=n}^N p\right) - \dtau \sum_{m=1}^M (m-1)\right)\right) \left(-n (t - t_4) - \dt \sum_{p=1}^{n-1} p \right) dt\right] \right].
\end{align}
Mathematica was used to evaluate this expression and arrive at (\ref{eq:dPhi_GG}).
\twocolumngrid

\bibliography{library}

\end{document}